	\documentclass[12pt, draftclsnofoot, onecolumn]{IEEEtran}
	\usepackage{pifont,bm,multicol,amsfonts,amsmath,color,amssymb,graphicx, epsfig,cite,psfrag,subfigure,algorithm,enumerate,stfloats,algorithmic,epstopdf,balance}

\begin{document}

\title{Deep Learning Optimized Sparse Antenna Activation for Reconfigurable Intelligent Surface Assisted Communication}

\author{Shunbo Zhang, Shun Zhang, \emph{Senior Member, IEEE}, Feifei Gao, \emph{Fellow, IEEE}, Jianpeng Ma, \emph{Member, IEEE}, Octavia A. Dobre, \emph{Fellow, IEEE}

\thanks{S. Zhang, S. Zhang and J. Ma are with the State Key Laboratory of Integrated Services Networks, Xidian University, Xi¡¯an 710071, P. R. China (e-mail: sbzhang$\_$19@stu.xidian.edu.cn, zhangshunsdu@xidian.edu.cn, jpmaxdu@gmail.com).}

    \thanks{F. Gao is with Department of Automation, Tsinghua University, State Key Lab of Intelligent Technologies and Systems, Tsinghua University, State Key for Information Science and Technology (TNList) Beijing 100084, P. R. China (e-mail: feifeigao@ieee.org).}


    \thanks{O. A. Dobre is with Faculty of Engineering and Applied Science, Memorial University of Newfoundland, St. John's NLAIC-5S7, Canada (e-mail: odobre@mun.ca).}

}

\maketitle

\vspace{-10mm}
\begin{abstract}

  To capture the communications gain of the massive radiating elements with low power cost,
  the conventional reconfigurable intelligent surface (RIS) usually works in passive mode.
  However, due to the cascaded channel structure and the lack of signal processing ability, it is difficult for RIS to obtain the individual channel state information and optimize the beamforming vector.
  In this paper, we add signal processing units for a few antennas at RIS to partially acquire the channels.
  To solve the crucial active antenna selection problem, we construct an active antenna selection network that utilizes the probabilistic sampling theory to select the optimal locations of these active antennas.
  With this active antenna selection network, we further design two deep learning (DL) based schemes, i.e., the channel extrapolation scheme and the beam searching scheme, to enable the RIS communication system.
  The former utilizes the selection network and a convolutional neural network to extrapolate the full channels from the partial channels received by the active RIS antennas, while the latter adopts a fully-connected neural network to achieve the direct mapping between the partial channels and the optimal beamforming vector with maximal transmission rate.
  Simulation results are provided to demonstrate the effectiveness of the designed DL-based schemes.
\end{abstract}

\maketitle
\thispagestyle{empty}
\vspace{-1mm}

\begin{IEEEkeywords}
	Deep learning, active RIS antenna elements, probabilistic sampling theory, channel extrapolation, beam searching
\end{IEEEkeywords}

\section{Introduction}
\label{introduce}

Recently, an emerging hardware technology called reconfigurable intelligent surface (RIS) has been considered as a promising technique for beyond 5G to capture the performance gain of the massive radiating elements \cite{Tang_RIS, Zhong_IRS1, Zhong_IRS3}.
The RIS consists of numerous reconfigurable reflecting elements, each of which is able to shift the phase of the incident electromagnetic waves by electronical controls \cite{RIS_Rui}.
With the equipped elements, an RIS can efficiently combine the reflected signals to achieve a high level of energy at the receiver side, and reconstruct the radio scattering environment into an intelligent one \cite{RIS_mag}.

Usually, the reflecting elements of RIS are working in passive mode, which leads to very low energy consumption \cite{passive1, passive2, passive3}.
Hence, RIS is easily integrated into the existing wireless systems \cite{Overview1}.
In \cite{Weighted_SumRate}, Guo \emph{et al.} proposed a low-complexity algorithm to jointly design the beamforming and the phase shifting at RIS elements to maximize the weighed sum-rate of all users.
In \cite{GMD}, Ying \emph{et al.} proposed a geometric mean decomposition-based beamforming for RIS-assisted millimeter wave hybrid multi-input-multi-output (MIMO) systems.
In \cite{IRS_THz}, Ning \emph{et al.} provided a hierarchical codebook design as the basis of beam training to reduce the complexity of channel estimation, and then proposed a cooperative channel estimation procedure for RIS-assisted system.
In order to gain the above advantages, accurate channel state information (CSI) is needed at RIS.
However, the shortage of RIS with full passive elements is that the channels from the source user to RIS and that from RIS to destination are coupled and cannot be separately estimated.

Generally, the objective of most existing RIS designs is to maximize the achievable rate at the user side by optimizing the beamforming vector \cite{achievable_rate1, achievable_rate2, Zhong_IRS2, achievable_rate3, achievable_rate4, achievable_rate5,Passive_Beam_Desi}.
In passive mode, RIS elements have no digital signal processing function.
Moreover, feeding back the achieved CSI at receiver to RIS for phase shifting may cost system bandwidth.
One solution is to place some baseband signal processing units at RIS and then directly estimate the desired channels, namely, some RIS elements could be activated during the communications process.
Then, there will be two stages, i.e., the channel estimation stage and the data transmission stage.
Obviously, the channel estimation stage would bring some extra power cost, but can simplify the signal processing of the data transmission.
In \cite{LIS1}, Jung \emph{et al.} utilized RIS with signal processing units to obtain the CSI by uplink pilot training and analyzed the performance of the system with a well-defined uplink frame structure and pilot contamination.
On the other hand, Alexandropoulos \emph{et al.} presented an RIS architecture comprising of passive elements, a simple controller, and a single radio frequency (RF) chain for baseband measurements \cite{hardware}.
Besides, they proposed an alternating optimization approach for explicit estimation of channel gains at RIS during dedicated training slots.

Obviously, all these works \cite{Weighted_SumRate}--\cite{hardware} are
closely dependent on hypothetical mathematical models. In the actual communication scenario,
the radio scattering conditions change rapidly with time \cite{CE_Ma, CE_Li}, which causes serious mismatch from the adopted mathematical model.
Deep learning (DL), aiming to dig certain performance gain from the data,
has undergone a renaissance with excellent performance and low complexity \cite{deep_RIS1, deep_RIS2, deep_RIS3, deep_CE1, deep_CE2, deep_CE3, deep_yang, deep_yang2}.
Hence, DL has been adopted to implement the signal processing tasks in RIS systems
and has achieved superior performance.
In \cite{DL_aid}, Khan \emph{et al.} proposed a DL method for channel estimation and phase angles in RIS-assisted wireless communication systems.
Gao \emph{et al.} developed an unsupervised learning based approach for passive beamforming in RIS-assisted communication systems \cite{unsupervised}.
In \cite{DRL}, Huang \emph{et al.} proposed a joint design of transmit beamforming and phase shifts based on deep reinforcement learning technique, which also has a standard formulation and low complexity.

However, to avoid large power consumption, the number of active elements on RIS should be limited.
Taha \emph{et al.} use randomly configured active elements to sub-sample the channels, and extrapolate the channels to all elements from those estimated at active elements \cite{LIS2}.
Moreover, they developed a DL-based solution to optimize the beamforming vector of RIS.
Obviously, the performance of the channel extrapolation is closely related to the selection of the activated RIS elements.
One commonly adopted way is to use uniform activation pattern as did in \cite{LIS2}.
However,
the best activation pattern should be related to the locations of users and RIS, together with the electromagnetic scattering environment, while
the uniform activation pattern may not be the optimal approach.

In this paper, we investigate the active element-aided RIS communication system and try to maximize the achievable rate for data transmission.
Specifically,
we add a few active elements at RIS and construct an active antenna selection network to find the optimal locations of these elements, where the probabilistic sampling theory is utilized to model the selection of the activated RIS elements as a continuous and differentiable function.
Furthermore, we design two schemes, i.e., the channel extrapolation scheme and
the beam searching schemes.
The former includes the active antenna selection network and a convolutional neural network (CNN) based channel extrapolation network that aims to extrapolates the full channels for
data transmission from the estimated partial channels,
while the latter adopts the active antenna selection network and a fully-connected neural network (FNN) based beam searching network that directly maps from the estimated partial channels to the optimal beamforming vector for data transmission.
Lastly, we design proper network off-line training to optimize both
the RIS activation pattern and the respective neural network (NN) of the two schemes.


The rest of this paper is organized as follows.
Section \ref{model} describes the system model and the problem formulation.
Section \ref{CES} introduces the DL-based channel extrapolation scheme.
Section \ref{BSS} presents the DL-based beam searching scheme.
Section \ref{simulation} provides the numerical results and conclusions are drawn in Section \ref{conclusion}.

Notations: Denote lowercase (uppercase) boldface as vector (matrix).
$(\cdot )^H $, $(\cdot )^T $, and $(\cdot )^{*} $ represent the Hermitian, transpose, and conjugate, respectively.
$\mathbb E \{\cdot \} $ is the expectation operator.
$\odot$ and $\otimes$ represent the Hadamard product operator and Kronecker product operator, respectively.
Denote $|\mathcal A | $ as the number of elements in set $\mathcal A$.
$[\mathbf A]_{i,j} $ and $[\mathbf A]_{\mathcal Q,:}$ (or $[\mathbf A]_{:, \mathcal Q} $) represent the $(i,j) $-th entry of $\mathbf A $ and the submatrix of $\mathbf A $ which contains the rows (or columns) with the index set $\mathcal Q $, respectively.
$[\mathbf B]_{:,:,i}$ is the $i$-th slice along the third dimension of a 3D matrix $\mathbf B$.
$v \sim \mathcal{CN} (0, \sigma^2)$ means that $v$ follows the complex Gaussian distribution with zero-mean and variance $\sigma^2$.
$\|\mathbf a\|$ is the $\ell_2$-norm of the vector $\mathbf a$.
The real and imaginary component of $x $ is expressed as $\Re(x)$ and $\Im(x)$, respectively.
Moreover, $\text{diag} (\mathbf x)$ is a diagonal matrix whose diagonal elements are formed with the elements of $\mathbf x$.

\section{System Model and Problem Formulation}
\label{model}

\subsection{RIS-assisted Communication System Model}

\begin{figure}
	\centering
	\includegraphics[width=100mm]{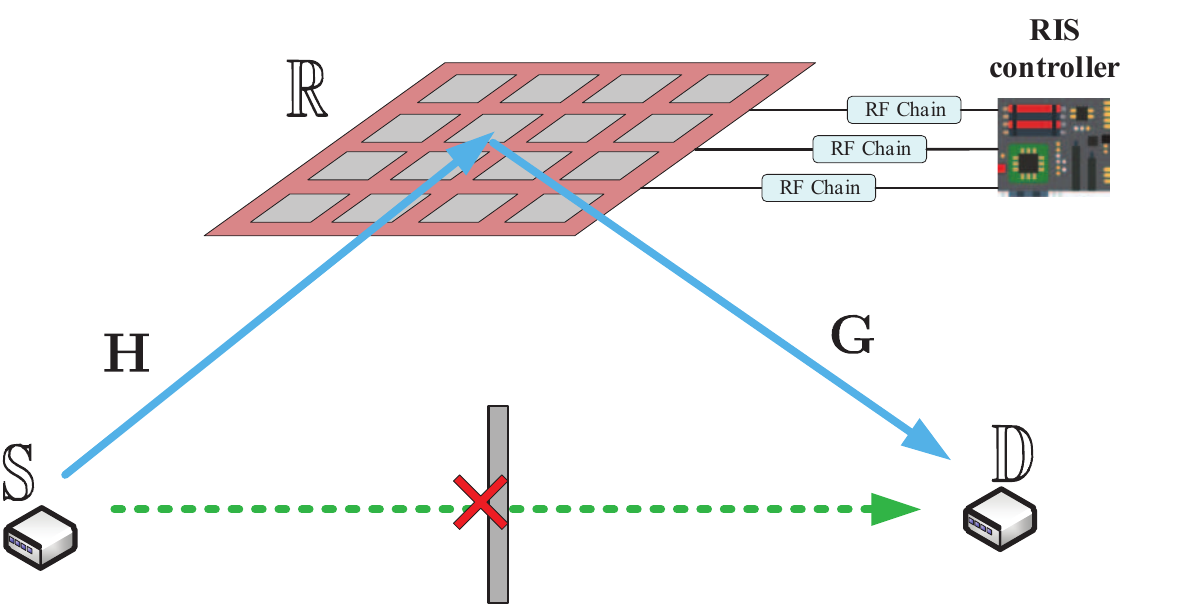}
	\caption{The RIS-assisted Communication System Model.}
	\label{system_model}
\end{figure}

Considering an RIS assisted communications system in Fig. \ref{system_model}, there are one single antenna transmitter $\mathbb S$, one single antenna receiver $\mathbb D$, and one RIS $\mathbb R$ with $N_v\times N_h=N$ reflecting elements in the form of a uniform planar array (UPA), where $N_v$ and $N_h$ are the number of elements in the vertical and horizontal dimension, respectively.
Denote the set of all reflecting elements at $\mathbb R$ as $\mathcal N$.
In particular, RIS can
assist the communications from $\mathbb S$ to $\mathbb D$ by reflecting the incident electromagnetic wave with tunable phase shift.

To combat the practical frequency selective fading, we adopt
orthogonal frequency division multiplexing (OFDM) scheme with $K$ subcarriers.
Generally,
the direct path between $\mathbb S$ and $\mathbb D$ tends to be blocked by the possible obstacles like
buildings and human bodies.
Hence, we mainly focus on the RIS assisted link.
Define the $\ell$-th tap of the time domain channel from $\mathbb S$ to $\mathbb R$ as \cite{CE_Liu}
\begin{align}
\label{time_domain_channel}
\breve{\mathbf{h}}_{\ell}
=\sum_{p=1}^{P_h}h_{p,f_c}
\delta(\ell T_s-\tau_{h,p})\mathbf{a}(\phi_{h,p},\varphi_{h,p}),
\end{align}
where $P_h$ is the number of the scattering paths along the link $\mathbb S\to \mathbb R$, $h_{p,f_c}$ is the equivalent complex channel gain of the $p$-th path at the carrier frequency $f_c$, $\tau_{h,p}$ is the time delay, $\delta({\cdot})$ denotes the Dirac function, $T_s$ is the system sampling period, and $\mathbf{a}(\phi_{h,p},\varphi_{h,p})\in\mathcal C^{N\times 1}$ represents the spatial steering vector of the RIS at the angles of arrival $\phi_{h,p}$, $\varphi_{h,p}$.
Then, the frequency domain channel vector at the $k$-th subcarrier from $\mathbb S$ to $\mathbb R$ can be derived as
\begin{align}
\label{frequency_domain_channel}
\mathbf{h}_{k}=[\mathbf{H}]_{:,k}
=\frac{1}{\sqrt{K}}\sum_{\ell=0}^{K-1}
\breve{\mathbf{h}}_{\ell}e^{-\jmath2\pi\frac{\ell k}{K}}
=\frac{1}{\sqrt{K}}\sum_{p=1}^{P_h}h_{p,f_c}e^{-\jmath2\pi\frac{k\tau_{h,p}}{KT_s}}\mathbf{a}(\phi_{h,p},\varphi_{h,p}),
\end{align}
where $\mathbf{H}=[\mathbf h_0,\mathbf h_1,\cdots,\mathbf h_{K-1}]\in\mathcal{C}^{N\times K}$ is the frequency domain channel matrix between $\mathbb S$ and $\mathbb R$.
The channel vector $\mathbf g_k\in\mathcal C^{N\times 1}$ at the $k$-th subcarrier from $\mathbb D$ to $\mathbb R$ can be similarly defined as \eqref{frequency_domain_channel}, and $\mathbf g_k^T$ is the channel from $\mathbb R$ to $\mathbb D$ by reciprocity.
Define $\mathbf G=[\mathbf g_0,\mathbf g_1,\cdots,\mathbf g_{K-1}]\in\mathcal C^{N\times K}$ as the frequency domain channel matrix between $\mathbb D$ and $\mathbb R$.

Then, the received signal of the $k$-th subcarrier at $\mathbb D$ can be written as
\begin{align}
\label{received_signal}
y_k=\mathbf{g}_{k}^T\bm{\Theta}
\mathbf{h}_{k}s_k+n_k
=(\mathbf{g}_{k}\odot\mathbf{h}_{k})^T
\bm{\theta}s_k+n_k,
\end{align}
where $s_k$ is the signal at the $k$-th subcarrier from $\mathbb S$,
$\bm{\Theta}=\text{diag}(e^{\jmath\theta_1},\cdots,
e^{\jmath\theta_N})\in\mathcal{C}^{N\times N}$ is a diagonal matrix
whose elements represent the phase shifts of RIS antennas,
and $n_k\sim\mathcal{CN}(0,\sigma^2)$ is the additive white Gaussian noise (AWGN) at $\mathbb D$.
Moreover, the diagonal elements in $\bm{\Theta}$ are collected
into the reflection beamforming vector $\bm{\theta}=[e^{\jmath\theta_1},\cdots,e^{\jmath\theta_N}]^T
\in\mathcal{C}^{N\times 1}$.

RIS can be reconfigured by a controller connected with RF chains,
and the corresponding phase shift is
set as a finite number of discrete values that belong to the quantized set
$\mathcal{A}\triangleq\left\{0,\Delta,\cdots,(2^b-1)\Delta\right\}$,
where $b$ is the number of quantization bits
and $\Delta=2\pi/2^b$ represents the quantization step size.
Since the considered RIS has $N$ digital phase shift elements, the reflection beamforming vector $\bm{\theta}$ would have $2^{bN}$ different choices.

\subsection{Problem Formulation}

The effective data are conveyed along the link $\mathbb S\rightarrow\mathbb R\rightarrow \mathbb D$ with center frequency $f_c$.
The aim is to maximize the achievable rate at $\mathbb D$ over all possible $\bm\theta$'s as
\begin{align}
\label{object}
(\text P1):~\max_{\bm{\theta}^{\star}}\ \ &R=\frac{1}{K}\sum_{k=0}^{K-1}\log_2\left(
1+\left|(\mathbf{g}_{k}
\odot\mathbf{h}_{k})^T
\bm{\theta}\right|^2/\sigma^2\right)\\
\label{constraint}
\text{s.t.}\ \ &\theta_n\in\mathcal{A}=\left\{0,\Delta,\cdots,
(2^b-1)\Delta\right\}, \forall n\in\mathcal{N}.
\end{align}

\begin{figure}
	\centering
	\includegraphics[width=120mm]{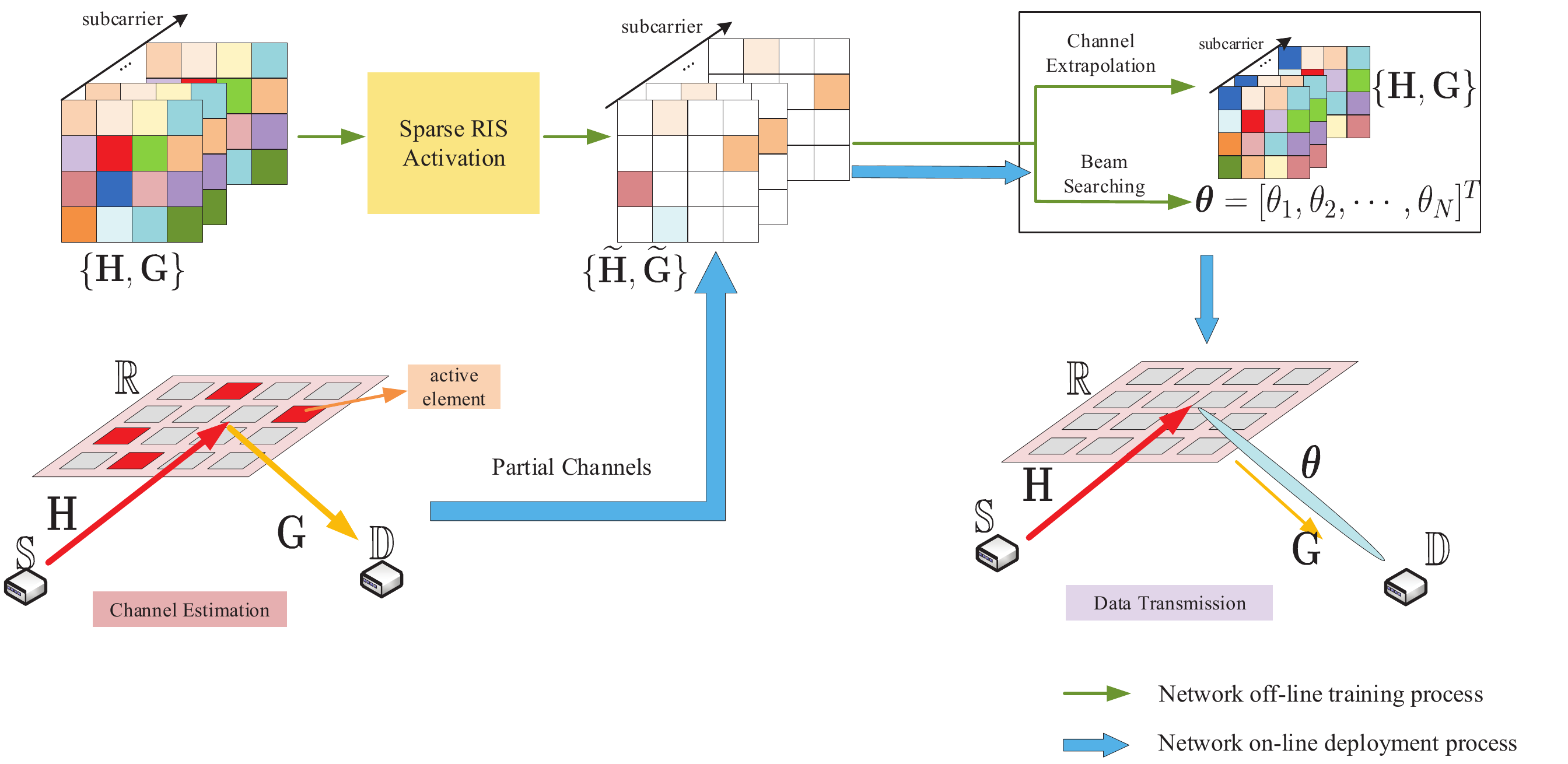}
	\caption{The architecture of the proposed schemes.}
	\label{framework}
\end{figure}

It can be checked that to find the optimal beamforming vector $\bm\theta^{\star}$, full channels $\mathbf H$ and $\mathbf G$ need to be acquired at $\mathbb R$, which is a hard task for traditional RIS due to the lack of signal processing function.
Intuitively, we can set a small part of the RIS elements, i.e., a set $\mathcal M$, as active sensors to obtain the partial channels between the transceivers and the RIS.
Then it is possible to further extrapolate the full channels from these partial ones, where $|\mathcal{M}|=M\ll N$ \cite{channel_map}.
Hence, how to design an efficient channel extrapolation scheme is the first challenge that needs to be solved.
Due to the limited number of RF chains and the requirement of low power cost, the number of active elements should be as small as possible, which may limit the performance of the channel extrapolation.
Thus, how to approach the maximal achievable rate in \eqref{object} by finding a quasi-optimal beamforming vector with fewer active elements is the second challenge that needs to be solved.
Note that the selection of the active antennas, i.e., the activation pattern $\mathcal M$, is crucial to overcoming the above two challenges but has not been solved to the best of the authors' knowledge.
In this paper, we design two different DL-based schemes, named channel extrapolation scheme and beam searching scheme, to separately address the aforementioned two challenges.
In both schemes, the activation pattern is optimized through the probabilistic sampling theory.
For both schemes, the corresponding off-line training and on-line deployment are introduced.
The unified architecture of these two schemes is shown in Fig. \ref{framework}.

\section{Deep Learning based Channel Extrapolation Scheme}
\label{CES}

\begin{figure}
	\centering
	\includegraphics[width=140mm]{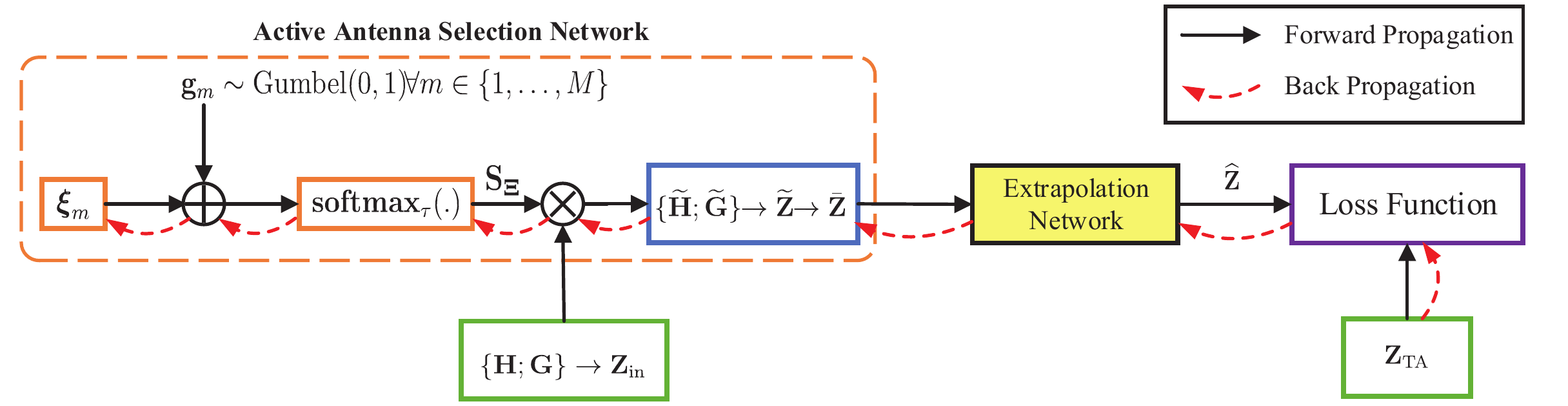}
	\caption{The structure of the designed DL-based channel extrapolation scheme.}
	\label{system architecture}
\end{figure}

As presented in Section \ref{model}, one main challenge for the RIS assisted communication system is the acquirement of the full channels $\mathbf H$ and $\mathbf G$ when designing the optimal beamforming vector $\bm\theta^{\star}$.
Traditionally, the receiver $\mathbb D$ can perform channel estimation to obtain the cascaded channel of $\mathbf H$ and $\mathbf G$ and feed back the CSI to $\mathbb R$.
However, the overheads of the channel estimation and the feedback would be high due to the massive number of reflecting elements on the RIS.
To overcome this bottleneck, we add signal processing function for a small part $\mathcal M$ of the RIS reflecting elements $\mathcal N$ and utilize these active antennas to acquire the individual channels between the transceivers and the RIS.
Define the channels between the active antennas $\mathcal M$ and the transceivers as two $M\times K$ matrices $\widetilde{\mathbf{H}}
=[\mathbf{H}]_{\mathcal{M},:}$ and $\widetilde{\mathbf{G}}
=[\mathbf{G}]_{\mathcal{M},:}$.
In this section, our first aim is to extrapolate the full channels from these partial channels, i.e., utilizing the partially known channels to recover the rest unknown channels between the transceivers and the rest antennas.
Note that the selection of the active antennas' locations, i.e, the activation pattern $\mathcal M$, would greatly affect the extrapolation performance.
Our second aim is to find an optimal activation pattern $\mathcal M$.
Since DL can effectively extract the latent and complex relationship among various datasets such as different channels, we design a DL-based channel extrapolation scheme to optimize both the activation pattern and the channel extrapolation ability.
The overall flow of the designed scheme is expressed as
\begin{align}
\label{mapping}
\{\mathbf{H},\mathbf{G}\} \xrightarrow{\text{spatial sub-sampling}}
\{\widetilde{\mathbf{H}},
\widetilde{\mathbf{G}}\} \xrightarrow{\text{channel extrapolation}}
\{\widehat{\mathbf{H}},\widehat{\mathbf{G}}\},
\end{align}
where $\widehat{\mathbf{H}}\in\mathcal C^{N\times K}$ and $\widehat{\mathbf{G}}\in\mathcal C^{N\times K}$ denote the recovery of $\mathbf H$ and $\mathbf G$, respectively.
The structure of \eqref{mapping} is shown in Fig. \ref{system architecture}, which contains the active antenna selection network and the channel extrapolation network.


\subsection{Active Antenna Selection Network}
\label{CES_selection}

Define the spatial compression ratio as $r=\frac{M}{N}$.
The effect of the active antenna selection network can be expressed as a sub-sampling matrix $\mathbf S$ on the full channel $\mathbf H$ and $\mathbf G$ as
\begin{align}
\label{subsampling1}
\widetilde{\mathbf H}=\mathbf S\mathbf H,\\
\widetilde{\mathbf G}=\mathbf S\mathbf G,
\end{align}
where $\mathbf{S}=[\mathbf s_0^T,\mathbf s_1^T,\ldots,\mathbf s_{M-1}^T]^T\in\{0,1\}^{M\times N}$ is a binary matrix and its $m$-th row vector $\mathbf s_m$ contains only one non-zero entry.

Generally, the Back Propagation (BP) algorithm is involved when training the DL model.
However, since the utilization of $\mathbf S$ is a discrete combination operation, it is difficult to define
the gradient differentiation.
To overcome this obstacle, we define a trainable matrix $\bm{\Xi}=[\bm{\xi}_0^T,\bm{\xi}_1^T,\cdots,\bm{\xi}_{M-1}^T]^T\in\mathcal R^{M\times N}$, whose $m$-th row vector is an independent categorical distribution $\bm\xi_m$ and the $n$-th entry of $\bm\xi_m$ is denoted as an unnormalized log-probability (logit) $\xi_{m,n}$.
Then, we leverage the probabilistic sampling strategy and replace $\mathbf S$ with a learned sub-sampling matrix $\mathbf S_{\bm\Xi}$, whose elements are controlled by $\bm\Xi$.

Since the DL model can only deal with real-valued number, we first separate the real and imaginary parts of $\mathbf{H}$ and $\mathbf{G}$, and then collect them into an $N\times K\times 4$ real-valued 3D matrix $\mathbf{Z}_{\text{in}}=[\Re(\mathbf{H});
\Im(\mathbf{H});
\Re(\mathbf{G});
\Im(\mathbf{G})]$ as the input data of the active antenna selection network.
The selection network implements the selection operation on $\mathbf H$ and $\mathbf G$ to obtain $\widetilde{\mathbf H}$ and $\widetilde{\mathbf G}$
by sub-sampling $\mathbf{Z}_{\text{in}}$ along its first dimension as
\begin{align}
\label{sampled_matrix}
\widetilde{\mathbf{Z}}
=[\Re(\mathbf{\widetilde{H}});
\Im(\mathbf{\widetilde{H}});
\Re(\mathbf{\widetilde{G}});
\Im(\mathbf{\widetilde{G}})]
=F_{\bm\Xi}(\mathbf{Z}_{\text{in}}),
\end{align}
where $\widetilde{\mathbf Z}\in\mathcal R^{M\times K\times 4}$ denotes the original output of the selection network, $F_{\bm\Xi}(\cdot)$ represents the sub-sampling function and is expressed as
\begin{align}
\label{subsampling2}
[\widetilde{\mathbf{Z}}]_{:,:,i}
=\mathbf{S}_{\bm{\Xi}}[\mathbf{Z}_{\text{in}}]_{:,:,i}, i=0,1,2,3.
\end{align}

Within the probabilistic sampling theory, $\mathbf s_m$ can be defined as \cite{subsampling}
\begin{align}
\label{onehot}
\mathbf{s}_m=\mathrm{one\_hot}(c_m),
\end{align}
where $\mathrm{one\_hot}(\cdot)$ denotes the one-hot encoding operation, $c_m\sim\text{Cat}(N,\bm\pi_m)$ is a categorical random variable with $\bm\pi_m=[\pi_{m,0},\pi_{m,1},\cdots,\pi_{m,N-1}]$ containing $N$ class probabilities.
For different categorical variables, i.e., $\forall m_1\neq m_2$, $c_{m_1}$ and $c_{m_2}$ are independent with each other.
Note that the result of $\mathrm{one\_hot}(c_m)$ is a $1\times N$ unit-vector and the index of the non-zero entry corresponds to the class of the drawn sample.
The larger $\pi_{m,n}$ means that the $m$-th rows of
$\widetilde{\mathbf H}$ and $\widetilde{\mathbf G}$
can be separately achieved from the $n$-th rows of $\mathbf H$ and
$\mathbf G$
with higher probability, namely, the $n$-th RIS element in $\mathcal N$ will be activated with higher priority.
Furthermore, we reparameterize $\pi_{m,n}$ with $\xi_{m,n}$ by using a $\mathrm{softmax}$ function as
\begin{align}
\label{softmax}
\pi_{m,n}=\frac{\exp(\xi_{m,n})}{\sum\limits_{n'=0}^{N-1}\exp (\xi_{m,n'})}.
\end{align}

In order to obtain an effective sample from the categorical distribution,
we resort to the $\mathrm{Gumbel}$-$\mathrm{Max}$ trick and generate a realization of $c_m$ as \cite{Statistical_Theory}
\begin{align}
c_m^{\prime}={\arg\max\limits_{n}\{\xi_{m,n}+w_{m,n}\}},
\end{align}
where $\{w_{m,0}, w_{m,1}, \cdots, w_{m,N-1}\}$ are independent and identically distributed samples following the $\mathrm{Gumbel}(0, 1)$ distribution. Correspondingly, $\mathbf s_m$ can be achieved from
$c_m^\prime$ as
\begin{align}
\mathbf s_m=\mathrm{one\_hot}\{\arg\max\limits_{n}\{\xi_{m,n}+w_{m,n}\}\}.
\end{align}

However, when conducting the above operation from $m=0$ to $M-1$,
the same row in
$\mathbf H$ and $\mathbf G$, i.e., the same RIS antenna element, may be repeatedly selected many times.
To avoid this case, we dynamically exclude the categories that have already been chosen.
Then, we renormalize the logits of the remaining categories and further implement
the $\mathrm{Gumbel}$-$\mathrm{Max}$ trick.

When training the selection network, $\boldsymbol\xi_m$ is iteratively updated
through the BP to complete the active antenna selection.
However, since the operator $\arg\max$ is non-differentiable, we resort to $\mathrm{softmax}_{\tau}$ function as a continuous and differentiable approximation of $\mathrm{one\_hot}\{\arg\max\}$.
Then, there is \cite{subsampling}
\begin{align}
\mathbf s_m=\lim_{\tau\rightarrow 0}\mathrm{softmax}_{\tau}(\boldsymbol{\xi}_m+\mathbf {w}_m)
=\lim_{\tau\rightarrow 0}\frac{\exp\{(\boldsymbol{\xi}_m+\mathbf {w}_m)/\tau\}}{\sum\limits_{n=0}^{N-1}
\exp\{(\xi_{m,n}+w_{m,n})/\tau\}},
\end{align}
where the temperature $\tau$ controls the softness of $\mathrm{softmax}_{\tau}$ and $\mathbf{w}_m=[w_{m,0},w_{m,1},\cdots,w_{m,N-1}]$ is the $1\times N$ $\mathrm{Gumbel}$ noise vector.
Note that lower $\tau$ means the generated Gumbel-Softmax distribution $\mathrm{softmax}_{\tau}(\boldsymbol{\xi}_m+\mathbf {w}_m)$ is closer to the categorical distribution.
During the selection network training, $\tau$ will be gradually reduced to approach the true discrete distribution.
The first-order partial derivative of $\mathbf s_m$ with respect to $\boldsymbol{\xi}_{m}$ can be written as
\begin{align}
\frac{\partial\mathbf{s}_m}{\partial\boldsymbol{\xi}_m^T}=
\frac{\partial}{\partial\boldsymbol{\xi}_m^T}
\mathbb{E}_{\mathbf{w}_m}\left[\mathrm{softmax}_{\tau}
(\boldsymbol{\xi}_m+\mathbf{w}_m)\right],~\tau>0.
\end{align}

In order to achieve a faster co-adaptation of the channel extrapolation network with different RIS activation patterns during training, we fill zeros into $\widetilde{\mathbf Z}$ after the sub-sampling operation and feed the data related with all RIS antenna elements into the following channel extrapolation network rather than those at the $M$ activated antennas.
Accordingly, the zero-filling operation on $\widetilde{\mathbf{Z}}$ is
\begin{align}
\label{zero_filling_1}
\bar{\mathbf{Z}}&=ZF(\widetilde{\mathbf{Z}}),
\end{align}
where $\bar{\mathbf{Z}}\in\mathcal R^{N\times K\times 4}$ represents the processed output of the selection network, the non-zero rows of $\bar{\mathbf{Z}}$ is consistent with $\widetilde{\mathbf{Z}}$
and their locations are the same with the original ones in $\mathbf{Z}_{\text{in}}$.

\subsection{Channel Extrapolation Network}
\label{CES_extrapolation}

\begin{figure}
	\centering
	\includegraphics[width=130mm]{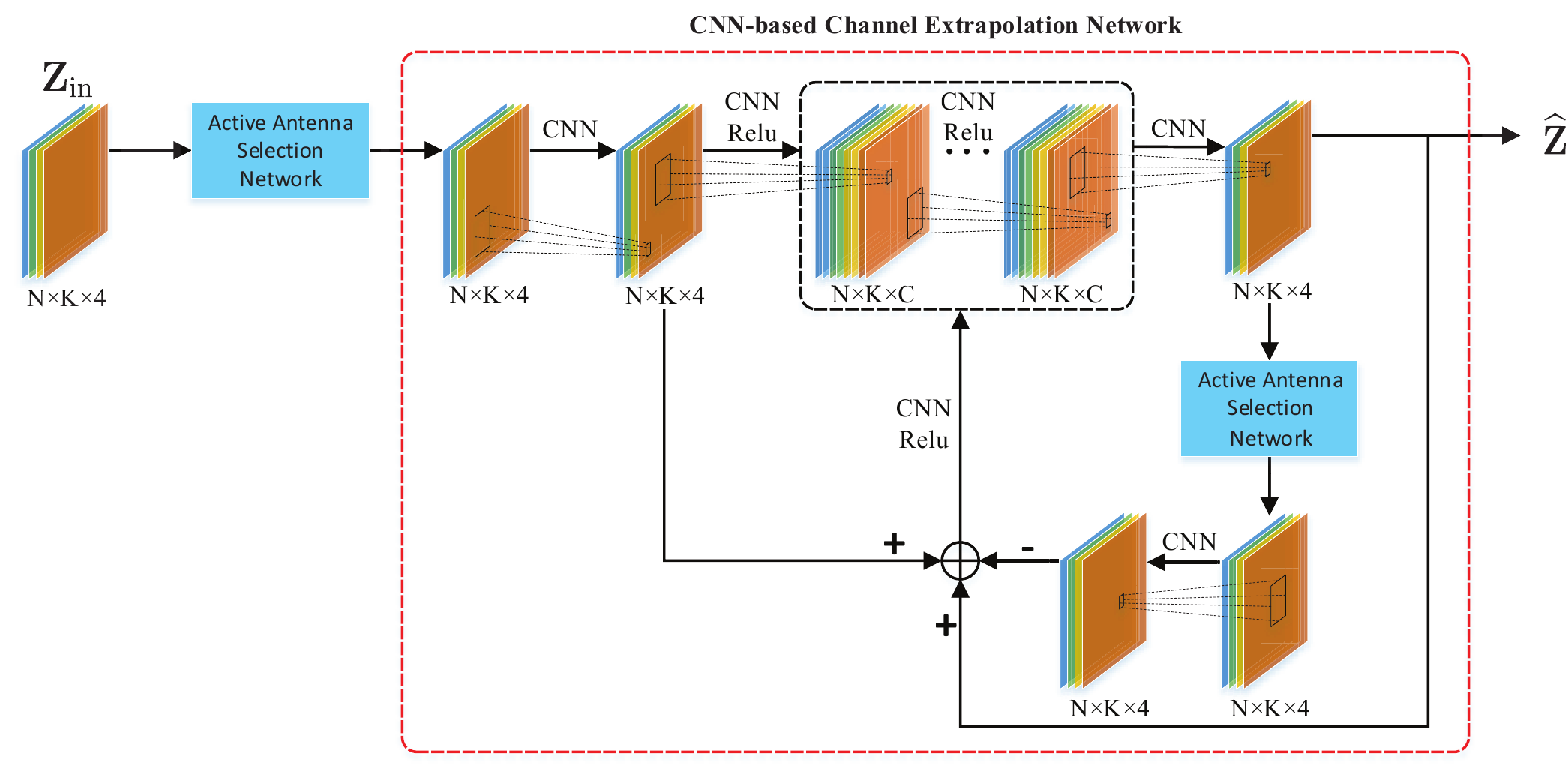}
	\caption{The structure of the proposed channel extrapolation network.}
	\label{cnn}
\end{figure}

The channel extrapolation network aims to simultaneously extrapolate the full channels $\mathbf{H}$ and $\mathbf{G}$ from the sub-sampled channels $\widetilde{\mathbf H}$ and $\widetilde{\mathbf G}$.
Since the channel matrices $\widetilde{\mathbf H}$, $\widetilde{\mathbf G}$, $\mathbf H$ and $\mathbf G$ are all 2D matrices, it is reasonable to utilize CNN to conducte the extrapolation task.
Moreover, to obtain an efficient extrapolation performance, we resort to the iterative proximal-gradient algorithm \cite{subsampling}, which is dedicated to solving the ill-posed linear measurement problem in \eqref{subsampling1}.
The structure of the CNN-based channel extrapolation network is shown in Fig. \ref{cnn}.
The output of the channel extrapolation network is expressed as
\begin{align}
\label{output1}
\widehat{\mathbf{Z}}
&=[\Re(\widehat{\mathbf{H}});
\Im(\widehat{\mathbf{H}});
\Re(\widehat{\mathbf G});
\Im(\widehat{\mathbf G})]
=G_{\mathcal{W}_C}(\bar{\mathbf{Z}}),
\end{align}
where $\widehat{\mathbf{Z}}\in\mathcal R^{N\times K\times 4}$ and the function $G_{\mathcal{W}_C}(\cdot)$ denotes the extrapolation operation learned by CNN, whose trainable parameters set is $\mathcal{W}_C$.

Consider $N_p$ proximal-gradient iteration and each iteration contains $N_q+2$ convolutional layers.
The total number of convolutional layers in the channel extrapolation network is $N_C=1+N_p\times(N_q+2)$.
In the $l$-th convolutional layer, the input 3D matrix is first zero-padded on each slice and is then processed by $N_l$ convolutional kernels of size $H\times W\times D_l$, where $H$, $W$ and $D_l$ represent the height, the width and the depth of the convolutional kernels, respectively.
Then, all $N_{l}$ convolutional kernels successively slide along the first two dimensions of the input to generate $N_{l}$ real-valued 2D feature matrices.
The size of each feature matrix depends on the height $H$ and width $W$ of the kernel, the number of zero-padding $P$ and the convolution stride $S$.
In our work, the hyperparameters like $P$ and $S$ should be designed carefully to ensure that the dimension of each slice remains unchanged after convolution.
The rectified linear unit ($\mathrm{ReLU}$) activation function is applied at the first $N_q$ convolutional layers in each iteration.
Finally, the output layer adopts 4 convolutional kernels to obtain the recovery version of the real and imaginary parts of the full channel matrices, as shown in (\ref{output1}).
More details about this part will be discussed in Section \ref{simulation}.
We present the feasibility of the channel extrapolation in Appendix \ref{appendix_a}.

After obtaining $\widehat{\mathbf{Z}}$, we further combine its real and imaginary parts to obtain the extrapolated full channels $\widehat{\mathbf H}$ and $\widehat{\mathbf G}$.

\subsection{Off-line Training for Active Antenna Selection and Channel Extrapolation Network}
\label{CES_training}

Define $\bm\omega_{l}$ as the vector containing trainable parameters of the $l$-th convolutional layer and  $\mathcal{W}_C=\{\bm\omega_{1},
\bm\omega_{2},\ldots,\bm\omega_{N_C}\}$ as the parameter set of the channel extrapolation network.
The set for off-line training is denoted as $\mathcal{D}$, where $|\mathcal{D}|=N_{tr}$ is the number of
off-line training samples.
Furthermore, a sample in $\mathcal D$ is an input-target pair written as $(\mathbf{Z}_{\text{in}}, \mathbf{Z}_{\text{TA}})$, where $\mathbf{Z}_{\text{TA}}$ is the extrapolation target and is equal to $\mathbf Z_{\text{in}}$ in our work since we need to acquire the original full channels.
During the off-line training phase, the tunable parameters in $\mathcal{W}_C$ and $\bm{\Xi}$ are trained by minimizing the
mean square error (MSE) between the output $\widehat{\mathbf{Z}}$ and
the target $\mathbf{Z}_{\text{TA}}$.
Thus, the loss function of the channel extrapolation network is written as
\begin{align}
\mathcal{L}_c
&=\frac{1}{4NKM_{tr}}\sum_{\mu=0}^{M_{tr}-1}\sum_{i=0}^3\Big\|
[\mathbf{Z}_{\text{TA}}^{\mu}]_{:,:,i}-
[\widehat{\mathbf{Z}}^{\mu}]_{:,:,i}\Big\|_F^2,
\end{align}
where $\|\mathbf{A}\|_F$ is the {$\mathrm{Frobenius}$ norm} of matrix $\mathbf{A}$ and $M_{tr}$ is the batch size for off-line training.

Besides, we promote the training of the parameters in $\bm\Xi$ towards one-hot distributions through penalizing convergence towards high entropy distribution for the active antenna selection network as
\begin{align}
\label{high_entropy}
\mathcal{L}_s=-\sum_{m=1}^{M}\sum_{n=1}^{N}\pi_{m,n}\log\pi_{m,n}.
\end{align}

When the parameters for the active antenna selection network and the channel extrapolation network are updated jointly, the resultant optimization problem of the channel extrapolation scheme is expressed as
\begin{align}
\left\{\widehat{\mathcal{W}}_C,\widehat{\bm{\Xi}}\right\}
=\arg\min_{\mathcal{W}_C,\bm{\Xi}}
(\mathcal{L}_c+\rho\mathcal{L}_s),
\end{align}
where the penalty multiplier $\rho$ evaluates the importance of different penalties.

During the off-line training phase, the adaptive moment estimation (Adam) \cite{stochastic_opt}
optimizer algorithm is adopted to achieve the optimal model
parameters $\widehat{\mathcal{W}}_C$ and $\widehat{\bm{\Xi}}$.
We use the learning rates $\eta_{\omega}$ and $\eta_{\xi}$ to respectively update the parameters in $\mathcal W_C$ and $\bm{\Xi}$, where $\eta_{\omega}< \eta_{\xi}$.
Moreover, we initially set $\tau$ as 5.0 and
gradually reduce it to 0.5 during training to approach the discrete distribution.
All elements in $\bm\Xi$ are initialized as $\xi_{m,n}\sim \mathcal{N}(0,0.05)$.

After completing the off-line training, the optimal activation pattern $\mathcal M$ can be obtained by extracting $\mathbf S_{\widehat{\bm\Xi}}$ from the trained active antenna selection network.
Note that $\mathbf S_{\widehat{\bm\Xi}}$ is controlled by the parameters in $\widehat{\bm\Xi}$ and the index of the non-zero entry in each row of $\mathbf S_{\widehat{\bm\Xi}}$ corresponds to the optimal location of an active antenna element on the RIS.
Moreover, the trained channel extrapolation network with parameters $\widehat{\mathcal W}_C$ can gain the ability to extrapolate the full channels from the given partial channels.

\subsection{On-line Deployment for Channel Extrapolation Network}
\label{CES_deployment}

\linespread{1.2}
\begin{algorithm}[!htp]
\caption{The training and deployment for channel extrapolation scheme}
\begin{algorithmic}[1]
\STATE \textbf{PHASE I:} Off-line training phase
\STATE \textbf{Require:} Training dataset $\mathcal{D}$, {the number} of iterations $N_{iter}$, $\tau_{start}=5,\tau_{end}=0.5$, and {the initialized} trainable parameters $\mathcal{W}_C$ and $\bm{\Xi}$.
\STATE Compute $\Delta\tau=\frac{\tau_{start}-\tau_{end}}{N_{iter}-1}$
\FOR{$i=1$ to $N_{iter}$ }
\STATE Draw mini-batches $\mathcal{D}_{m}$: a random subset of $\mathcal{D}$
\STATE Draw $\mathrm{Gumbel}$ noise vectors $\mathbf{w}_m$ for $m\in\{0,...,M-1\}$
\STATE Compute $\mathbf{S}_{\bm{\Xi}}=[\mathbf s_0^T,\mathbf s_1^T,\ldots,\mathbf s_{M-1}^T]^T$ using
$\mathbf s_m=\mathrm{one\_hot}\{\arg\max\limits_{n}\{\xi_{m,n}+w_{m,n}\}\}$
for $m\in\{0,...,M-1\}$, and dynamically exclude the repeatedly selected elements
\STATE Sub-sample the input as $[\widetilde{\mathbf{Z}}]_{:,:,i}
=\mathbf{S}_{\bm{\Xi}}[\mathbf{Z}_{\text{in}}]_{:,:,i}$ for $i=0,1,2,3$
\STATE Achieve the input of CNN-based channel extrapolation network as $\bar{\mathbf{Z}}=ZF(\widetilde{\mathbf{Z}})$
\STATE Compute the output of CNN-based channel extrapolation network as $\widehat{\mathbf{Z}}
=G_{{\mathcal W_C}}(\bar{\mathbf{Z}})$
\STATE Compute the  loss function as $\mathcal{L}_c+\rho\mathcal{L}_s$
\STATE Set $\tau=\tau_{start}-(i-1)\Delta \tau$
\STATE Update $\frac{\partial}{\partial\bm{\xi}_m^T}\mathbb E_{\mathbf{w}_m}\left[\mathrm{softmax}_{\tau}
(\bm{\xi}_m+\mathbf{w}_m)\right],~\tau>0$
\STATE Use the Adam optimizer to update $\mathcal{W}_C$ and $\bm{\Xi}$
\ENDFOR
\STATE Acquire the learned logit matrix $\widehat{\bm{\Xi}}$, the optimal activation pattern $\mathcal M$ and channel extrapolation network parameters $\widehat{\mathcal{W}}_C$
\STATE Determine the locations of the $M$ active antenna elements on $\mathbb R$ with $\mathcal M$
\STATE \textbf{PHASE II:} On-line deployment phase
\STATE $\mathbb R$ utilizes the $M$ active antenna elements as channel sensors to obtain $\widetilde{\mathbf{H}}$ and $\widetilde{\mathbf{G}}$
\STATE Feed $\widetilde{\mathbf{H}}$ and $\widetilde{\mathbf{G}}$ into
 the trained channel extrapolation network to obtain $\widehat{\mathbf{H}}$ and $\widehat{\mathbf{G}}$
\STATE Solve the projection problem in  \eqref{beam_search_equ_pro} to  periodically obtain the reflection beamforming vector $\bm{\theta}^o$ on $\mathbb R$
\STATE Adopt $\bm{\theta}^o$ to implement  the communication along the link $\mathbb S\rightarrow\mathbb R\rightarrow\mathbb D$
\end{algorithmic}
\end{algorithm}

As described in (P1), our next step is to calculate the optimal reflection bemforming vector $\boldsymbol\theta^o$ with the recovered $\widehat{\mathbf H}$ and $\widehat{\mathbf G}$ in the on-line phase.

During the channel estimation stage, the $M$ active antenna elements act as channels sensors to obtain the partial channels $\widetilde{\mathbf H}$ and $\widetilde{\mathbf G}$ by standard approaches such as least square (LS) estimation and minimum-mean square error (MMSE) estimation.
Subsequently, the trained channel extrapolation network can rapidly output the extrapolated channels $\widehat{\mathbf{H}}$ and $\widehat{\mathbf{G}}$ with fed $\widetilde{\mathbf{H}}$ and $\widetilde{\mathbf{G}}$.
Obviously, with the monotonicity of the logarithmic function and the independence of noise, (P1) can be equivalently transformed as
\begin{align}
\label{beam_search_equ}
\text{(P2)}:~\bm{\theta}^{\star}&=\arg\max_{\bm{\theta}}\sum_{k=1}^{K}
\left(\left|\left([\widehat{\mathbf{G}}]_{:,k}
\odot[\widehat{\mathbf{H}}]_{:,k}\right)^T\bm{\theta}\right|^2
\right)\notag\\
\text{s.t.}\ \ \theta_n&\in\mathcal{A}=\left\{0,\Delta,\cdots,
(2^b-1)\Delta\right\}, \forall n\in\mathcal{N}.
\end{align}

Note that without the constraint, the optimal solution can be readily obtained as
\begin{align}
\label{optimal_solution}
\bm{\theta}^{\star} = \frac{\sum_{k=1}^{K}\left([\widehat{\mathbf{G}}]_{:,k}
\odot[\widehat{\mathbf{H}}]_{:,k}\right)^\ast}{\|\sum_{k=1}^{K}\left([\widehat{\mathbf{G}}]_{:,k}
\odot[\widehat{\mathbf{H}}]_{:,k}\right)^\ast\|}.
\end{align}

Then, with a pre-defined set of the phase shift range $ \mathcal A $ for $ \boldsymbol \theta $, we can achieve the solution $ \boldsymbol \theta^o$  by solving the following projection problem
\begin{align}
\label{beam_search_equ_pro}
\bm{\theta}^o=&\arg\min_{{\theta}_n\in\mathcal A}||\bm{\theta}^{\star} - \bm{\theta}||^2.\end{align}

Within the subsequent data transmission stage, $\mathbb R$ utilizes $\bm{\theta}^o$ to assist the communication between $\mathbb S$ and $\mathbb D$.
For clarity, we present the details about the off-line training and the on-line deployment of the channel extrapolation scheme in Algorithm 1.

\section{Deep Learning based Beam Searching Scheme}
\label{BSS}

To overcome the second challenge presented in Section \ref{model}, we need to further reduce the number of active antenna elements compared with the channel extrapolation scheme and approach the the maximal achievable rate for data transmission.
We first adopt a pre-define codebook $\mathcal{B}$ which contains the candidate reflection beamforming vector $\bm\theta$ and is in the same order of the number of the RIS reflecting elements.
The more detailed design of the codebook $\mathcal B$ will be presented in Section \ref{simulation}.
It is worth noting that the codebook $\mathcal B$ is a suboptimal option compared with the quantized set $\mathcal A$ but can reduce the training overhead.
Then, if the number of possible solutions for $\bm{\theta}$ is limited and is not too large, some coarse partial channels $\widetilde{\mathbf{H}}$ and $\widetilde{\mathbf{G}}$ that are obtained from fewer active antennas can be utilized to establish a well mapping
between these partial channels and an optimal beamforming vector $\bm{\theta}^s$ in $\mathcal{B}$.
Note that the selection of activation pattern $\mathcal M$ would also greatly affect the performance of the mapping.
Hence, we further propose a DL-based beam searching scheme to optimize the activation pattern $\mathcal M$ and to extract the hidden relationship between the partial channels and the optimal beamforming vector in $\mathcal B$.
The overall flow of the designed scheme is expressed as
\begin{align}
\label{mapping2}
\{\mathbf{H},\mathbf{G}\} \xrightarrow{\text{spatial sub-sampling}}
\{\widetilde{\mathbf{H}},
\widetilde{\mathbf{G}}\} \xrightarrow{\text{beam searching}}
\{\bm\theta^s\}.
\end{align}

Similar to the channel extrapolation scheme, the beam searching scheme also contains two main parts, i.e., the active antenna selection network and the beam searching network, to separately conduct the spatial sub-sampling operation and the beam searching operation in \eqref{mapping2}.

\subsection{Active Antenna Selection Network}
\label{BSS_selection}

Since the structure of the active antenna selection network in the beam searching scheme is similar to that in the channel extrapolation scheme in Fig. \ref{system architecture}, we omit the description of this part and directly propose the beam searching network in the following part.

\subsection{Beam Searching Network}
\label{BSS_searching}

\begin{figure}
	\centering
	\includegraphics[width=140mm]{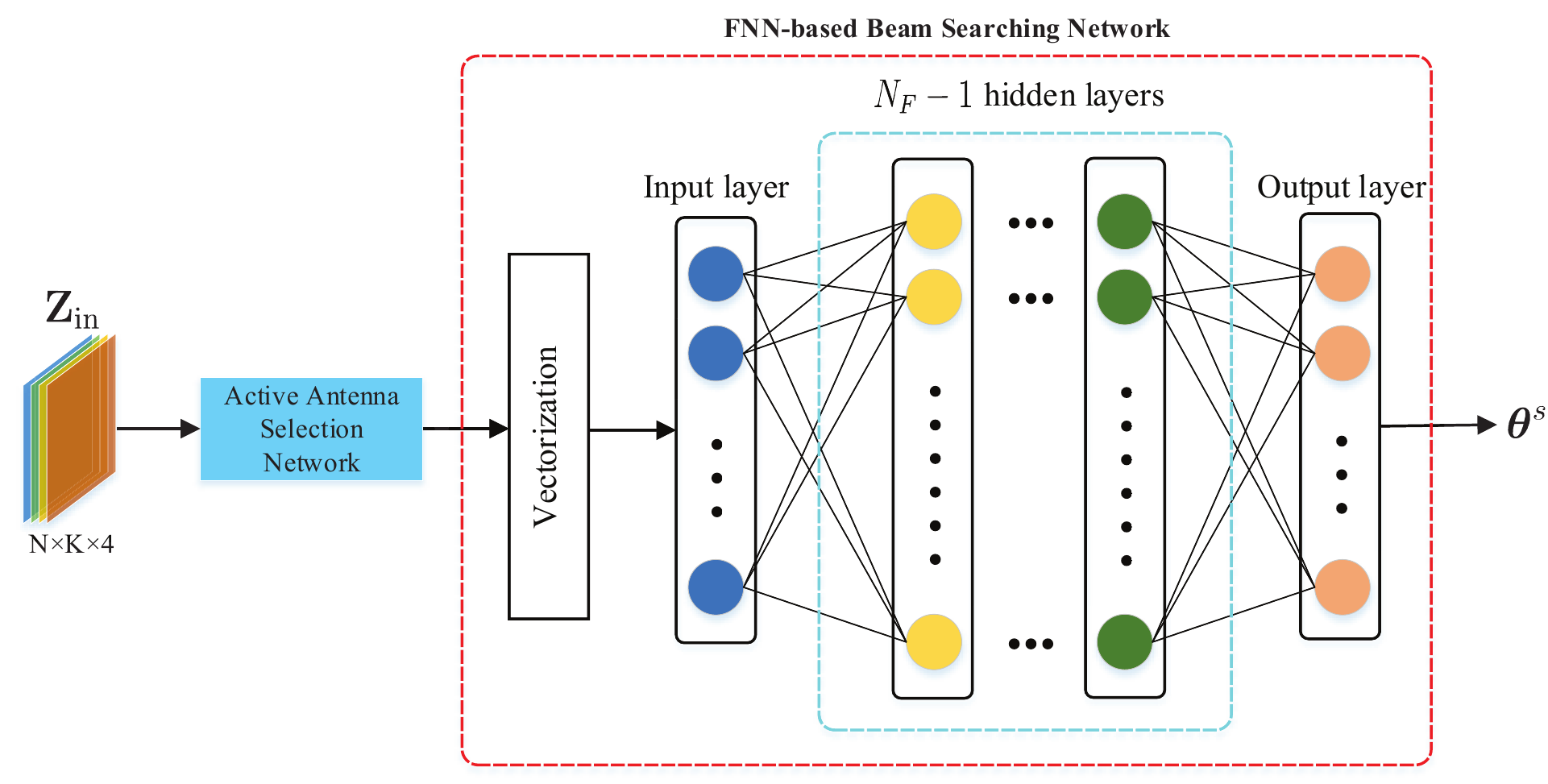}
	\caption{The structure of the proposed beam searching network.}
	\label{fnn}
\end{figure}

The beam searching network aims to find an optimal beamforming vector $\bm{\theta}^{s}$ in the codebook $\mathcal B$ with given $\widetilde{\mathbf H}$ and $\widetilde{\mathbf G}$.
Compared with the CNN-based channel extrapolation network which requires relatively expensive training overhead due to the high-dimensional output, the output dimension of the beam searching network is much lower.
This inspires us to adopt FNN for the beam searching network to find out the optimal beamforming vector $\bm\theta^s$, as shown in Fig. \ref{fnn}.
Use the codebook $\mathcal{B}$ to construct the training target so that once the beam searching network is fed with an input $\bar{\mathbf Z}$, it can pick up one $\bm{\theta}^s$ from $\mathcal{B}$ to maximize the achievable rate $R$.
Accordingly, the selection of $\boldsymbol\theta^s$ can be converted to a multi-classification problem.
In other words, since the training of the active antenna selection network and the beam searching network are implemented jointly, the corresponding index for $\bm{\theta}^s$ in $\mathcal{B}$ can be viewed as a label attached to a specific full channel pair $\{\mathbf H,\mathbf G\}$, i.e., the input data $\mathbf Z_{\text{in}}$ of the active antenna selection network in the beam searching scheme.
Hence, the expected output of the beam searching network can be transformed into $\mathbf{p}^s$, which is a one-hot encoding vector of size $|\mathcal{B}|\times 1$ and the index of the non-zero element in {$\mathbf p^s$}
indicates the location of $\bm{\theta}^s$ in $\mathcal B$.

We adopt the FNN-based beam searching network to find out the optimal beamforming vector $\boldsymbol\theta^s$ as
\begin{align}
\widehat{\mathbf{p}^s}
=G_{\mathcal{W}_B}(\bar{\mathbf{Z}}),
\end{align}
where $\mathcal{W}_B$ represents the trainable parameters set of the FNN-based beam searching network and $\widehat{\mathbf{p}^s}$
represents the output of the beam searching network.

In the beam searching network, the designed FNN contains one input layer, $N_F-1$ hidden layers and one output layer, as shown in Fig. \ref{fnn}.
The input $\bar{\mathbf Z}$ is first flatten into a vector to the input layer and is then fully-connected with the subsequent hidden layer.
The $l$-th hidden layer contains $N_l$ output nodes and the $\mathrm{Leaky~ReLUs}$ is used as the activation functions for all the hidden layers.
Moreover, $\mathrm{dropout}$ \cite{dropout} is utilized for all but the last hidden layer to avoid overfitting.
Finally, the output layer adopts a $\mathrm{softmax}$ function as defined in \eqref{softmax} to obtain the output $\widehat{\mathbf{p}^s}$, which includes the corresponding probabilities for all the classifications with respect to
the input data $\mathbf Z_{\text{in}}$.
Then, the index of the maximum value in $\widehat{\mathbf{p}^s}$ is the estimated location of $\bm\theta^s$ in $\mathcal B$.
We present the feasibility of the beam searching in Appendix \ref{appendix_b}.

\subsection{Training and Deployment for Active Antenna Selection and Beam Searching Network}
\label{BSS_learning}

During the off-line training phase, we use the available full channel data pair $\{\mathbf{H},\mathbf G\}$ and the per-defined codebook $\mathcal B$ to get the corresponding label $\mathbf{p}^s$.
Accordingly, a sample in the training set $\mathcal{D}$ is denoted by $(\mathbf{Z}_{\text{in}},\mathbf{p}^s)$, where the input data $\mathbf Z_{\text{in}}$ for the beam searching scheme has the same definition as that for the channel extrapolation scheme in \eqref{sampled_matrix}.
Furthermore, we use the categorical cross entropy between the output and label as the loss function of the beam searching network, which can be expressed as
\begin{align}
\mathcal{L}_b=-\frac{1}{M_{tr}}\sum_{\mu=0}^{M_{tr}-1}
\sum_{i=0}^{|\mathcal{B}|-1}p_i^{\mu}\log\widehat{p}_i^{\mu},
\end{align}
where $p_i$ and $\widehat{p}_i$ are separately the $i$-th element in $\mathbf{p}^s$ and $\widehat{\mathbf{p}^s}$, and $M_{tr}$ is the batch size for network training.

Considering the loss of the active antenna selection network in (\ref{high_entropy}), the resultant optimization problem for the beam searching scheme is denoted as
\begin{align}
\left\{\widehat{\mathcal{W}}_B,\widehat{\bm{\Xi}}\right\}
=\arg\min_{\mathcal{W}_B,\bm{\Xi}}(\mathcal{L}_b
+\rho\mathcal{L}_s).
\end{align}

During the off-line training phase, the Adam optimizer is adopted to achieve the optimal model
parameters $\widehat{\mathcal{W}}_B$ and $\widehat{\bm{\Xi}}$.
Since the off-line training for the beam searching network is similar with that for the channel extrapolation network in Section \ref{CES_training}, we omit some description due to space limitation.
More details about the training of the beam searching network will be specified in Section \ref{simulation}.

After completing the off-line training, the optimal activation pattern $\mathcal M$ can be acquired from the trained active antenna selection network, and the optimal locations of the active antenna elements are determined.

In the following deployment phase, once $\mathbb R$ obtains the partial channels $\widetilde{\mathbf H}$ and $\widetilde{\mathbf G}$ from the channel estimation stage, it can directly determine the optimal beamforming vector $\bm{\theta}^s$ in the codebook $\mathcal B$ to assist the communication between $\mathbb S$ and $\mathbb D$ during the subsequent data transmission stage.


\section{Simulation Results}
\label{simulation}
In this section, we evaluate the performance of the designed channel extrapolation scheme and beam searching scheme through numerical simulations.

\subsection{Communication Scenario and DeepMIMO Dataset}

Considering the presented RIS-aided communication system described in Section \ref{model}, it is reasonable to adopt a realistic electromagnetic environment to generate the channels. Therefore, we resort to the indoor massive MIMO scenario `I1' of the DeepMIMO dataset \cite{DeepMIMO}, which is generated based on the Wireless InSite \cite{wireless_insite} and is widely used in DL applications for massive MIMO systems.

Correspondingly, the primary parameters for the simulation are listed in TABLE~I.
Adopt the BS 8 in the `I1' scenario as the RIS of the system model.
The RIS is set as an UPA with $8\times 8$ ($N=64$) antennas as its elements.
In FDD model, the forward link $\mathbb S\rightarrow\mathbb R\rightarrow \mathbb D$ and the backward link $\mathbb D\rightarrow\mathbb R\rightarrow \mathbb S$ work in different frequency bands.
Intuitively, different activation patterns $\mathcal M$ could be selected at different frequency bands for the forward and the backward links, respectively.
However, this solution would increase the system's power consumption and decrease its spectral efficiency.
One feasible method is that the two links share the same activated RIS elements from one frequency band.
Thus, there exists frequency mismatch between the estimated channels and the channels to be extrapolated.
To exhibit the ability of channel extrapolation between different frequencies, we set the carrier frequencies for the channel estimation stage and the data transmission stage as $f_a=2.4$ GHz and $f_c=2.5$ GHz, respectively.
Denote $\mathbf H^a$ and $\mathbf G^a$ as the channel matrices with carrier frequency $f_a$.
For UPA, the antenna spacing $d$ is set to $\frac{\lambda_c}{2}$ and $\frac{\lambda_c}{4}$ for comparison.
Moreover, we select the users located within the regions from the 1st row to the 200th row and from the 201th row to the 400th row in the `I1' scenario as the transmitters $\mathbb S$ and receivers $\mathbb D$, respectively.
Since each row in the aforementioned regions contains 201 users, the total number of users is 80400.
We select each $\mathbb S$-$\mathbb D$ pair one-to-one from their corresponding regions to further generate 40200 samples.
The bandwidth of the OFDM system is set as 100 MHz, while the number of sub-carriers is set as $K=64$.
The channels $\mathbf H^a$, $\mathbf G^a$, $\mathbf H$ and $\mathbf G$ are generated from the DeepMIMO dataset generation code \cite{DeepMIMO}.
Typically, we adopt $\mathbf C_{N_v,r_1}\otimes \mathbf C_{N_h,r_2}$ as the beamforming codebook $\mathcal B$ to match the structure of the proposed RIS, where $\mathbf C_{N_v,r_1}\in\mathcal C^{N_v\times r_1N_v}$ and $\mathbf C_{N_h,r_2}\in\mathcal C^{N_h\times r_2N_h}$ are separately the beamforming codebooks along the vertical and horizontal dimension, $r_1$ and $r_2$ are the over-sampling coefficients for $\mathbf C_{N_v,r_1}$ and $\mathbf C_{N_h,r_2}$, respectively.
The $(i,j)$-th entry of $\mathbf C_{N_v,r_1}$ is defined as $\mathbf [\mathbf C_{N_v,r_1}]_{i,j}=\frac{1}{\sqrt{N_v}}e^{-\jmath2\pi\frac{d}{\lambda_c}
i\cos(\frac{\pi}{r_1N_v}j)}, i=0,1,\cdots,N_v-1, j=0,1,\cdots,r_1N_v-1$ and the entries in $\mathbf C_{N_h,r_2}$ have the similar definition.


\begin{table}[!t]
		\centering
		\renewcommand{\arraystretch}{1.3 }
		\caption{The adopted DeepMIMO dataset parameters.}
		\label{table}
		\begin{tabular}{c c  }
			\hline
            Parameter &value\\
            \hline
            \hline
			Name of scenario &I1\\
			\hline
			The carrier frequency of channel estimation and data transmission & 2.4 GHz, 2.5 GHz\\
			\hline
			Number of BS antennas in (x, y, z) & (8, 8, 1) \\
			\hline
			Number of paths &  5\\
			\hline
			Active users as the transmitters $\mathbb S$& Row 1 to 200 \\	
			\hline
		    Active users as the transmitters $\mathbb D$ &  Row 201 to 400\\
			\hline
			System bandwidth& 100 MHz\\	
			\hline
		    Number of OFDM sub-carriers&  64\\
			\hline
			\hline	
		\end{tabular}	
	\end{table}

\subsection{Network Parameters Configuration}
In the channel extrapolation scheme, one sample of the dataset is composed of two channel sets $\{\mathbf H^a,\mathbf G^a\}$ and $\{\mathbf H,\mathbf G\}$.
Employ $80\%$ of the dataset for network training and the rest for testing.
Considering the CNN-based channel extrapolation network in Section \ref{CES_extrapolation}, we use $3\times 3$ convolutional kernels and set $P=1$ and $S=1$ for all the convolutional layers.
Specially, we set $N_p=5$ and $N_q=6$ to learn a powerful proximal operator as shown in Fig. \ref{cnn}.
The initial parameters for the learning rates are set as $\eta_{\xi}=1e-3$ and $\eta_{\omega}=1e-4$, respectively, and the penalty multiplier $\rho$ is taken as $1e-4$.
The Adam optimizer is used for the network training with batch size 16.
We conduct the training of the active antenna selection network and the channel extrapolation network until the training loss converges.
TABLE~II provides the layer parameters of the channel extrapolation network.

\begin{table}[!t]
		\centering
		\renewcommand{\arraystretch}{1.3 }
		\caption{Layer Parameters for the CNN-based Channel Extrapolation Network.}
		\label{table}
		\begin{tabular}{c c c c c c  }
			\hline
			Layer &Output size &Initialization &Activation &Kernel size &Strides\\
			\hline
            \hline
			$1\times$ Conv2D &$64\times 64\times 4$ &Glorot uniform &None &$3\times 3$ &$1\times 1$\\
			\hline
			$6\times$ Conv2D (proximal-gradient iteration)&$64\times 64\times 64$ &Glorot uniform &ReLU &$3\times 3$ &$1\times 1$\\
			\hline
			$1\times$ Conv2D (proximal-gradient iteration)&$64\times 64\times 4$ &Glorot uniform &None &$3\times 3$ &$1\times 1$\\
			\hline
			$1\times$ Conv2D (proximal-gradient iteration)&$64\times 64\times 4$ &Glorot uniform &None &$3\times 3$ &$1\times 1$\\	
			\hline
            \hline	
		\end{tabular}	
	\end{table}

\begin{table}[!t]
		\centering
		\renewcommand{\arraystretch}{1.3 }
		\caption{Layer Parameters for the FNN-based Beam Searching Network.}
		\label{table}
		\begin{tabular}{c c c c c c  }
			\hline
			Layer &Output size &Initialization &Activation\\
			\hline
            \hline
			Flatten &16384 &- &-\\
			\hline
			FNN 1&16384 &Glorot uniform &Leaky ReLU ($\alpha=0.2$)\\
			\hline
			Dropout 1 (50\%)&16384 &- &-\\
            \hline
			FNN 2&4096 &Glorot uniform &Leaky ReLU ($\alpha=0.2$)\\
			\hline
			Dropout 2 (50\%)&4096 &- &-\\
            \hline
			FNN 3&4096 &Glorot uniform &Leaky ReLU ($\alpha=0.2$)\\
			\hline
			Dropout 3 (50\%)&4096 &- &-\\
            \hline
			FNN 4&2048 &Glorot uniform &Leaky ReLU ($\alpha=0.2$)\\
			\hline
			FNN 5&256 &Glorot uniform &Softmax\\
			\hline
            \hline	
		\end{tabular}	
	\end{table}

\begin{figure}[htp]
	\centering
	\includegraphics[width=105mm]{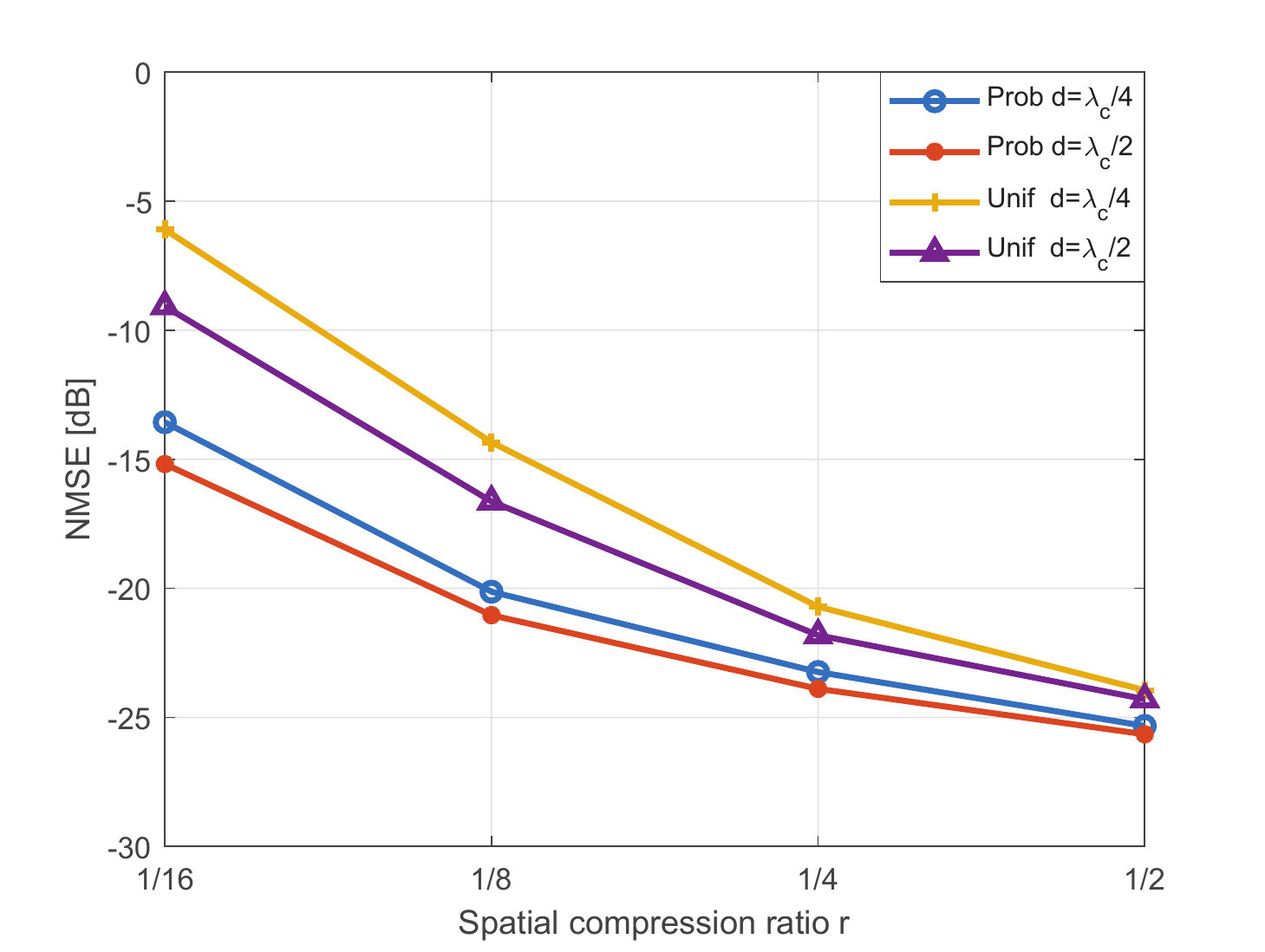}
	\caption{The NMSEs of channel extrapolation versus the spatial compression ratio $r$.}
	\label{re_compress_ratio}
\end{figure}

With respect to the beam searching scheme, the input data is the same as that for the channel extrapolation scheme.
We set the over-sampling coefficients for the codebook $\mathcal B$ as $r_1=r_2=2$.
Each label $\mathbf p^s$ in a sample of the training dataset corresponds to an optimal beamforming vector $\bm\theta^s$ searched in $\mathcal B$.
The dataset is divided with the same ratio as that for the channel extrapolation scheme.
Adopt four hidden layers for the FNN-based beam searching network in Section \ref{BSS_searching}.
The initial learning rates are $\eta_{\xi}=1e-2$ and $\eta_{\omega}=1e-4$, respectively, and the penalty multiplier $\rho$ is $1e-4$.
The Adam optimizer is used for the network training with batch size 256 and the epoch for training is set to 500.
The layer parameters of the beam searching network are listed in TABLE~III.



\subsection{Performance Evaluation}

Fig. \ref{re_compress_ratio} depicts the normalized MSEs (NMSEs) for the channel extrapolation scheme versus the spatial compression ratio $r$.
Note that the curves labeled by `Unif' correspond to the active antenna selection network with uniform selection strategy, while the ones marked by `Prob' represent the performance of the active antenna selection network with the proposed probabilistic selection strategy.

It can be checked that all the NMSE curves decrease with the increase of $r$, where $r\in\{\frac{1}{16},\frac{1}{8},\frac{1}{4},\frac{1}{2}\}$.
Moreover, it can be found that the performance of the proposed channel extrapolation scheme with probabilistic selection strategy is better than that with uniform selection strategy for both $d=\frac{\lambda_c}{4}$ and $d=\frac{\lambda_c}{2}$.
Specially, compared with the case of $d=\frac{\lambda_c}{4}$, the proposed scheme with $d=\frac{\lambda_c}{2}$ can obtain a better performance.
This is because that a smaller antenna spacing leads to a higher correlation between the channels of neighboring antennas that can not be distinguished, which damages the CNN's extrapolation performance.

\begin{figure}
	\centering
	\includegraphics[width=105mm]{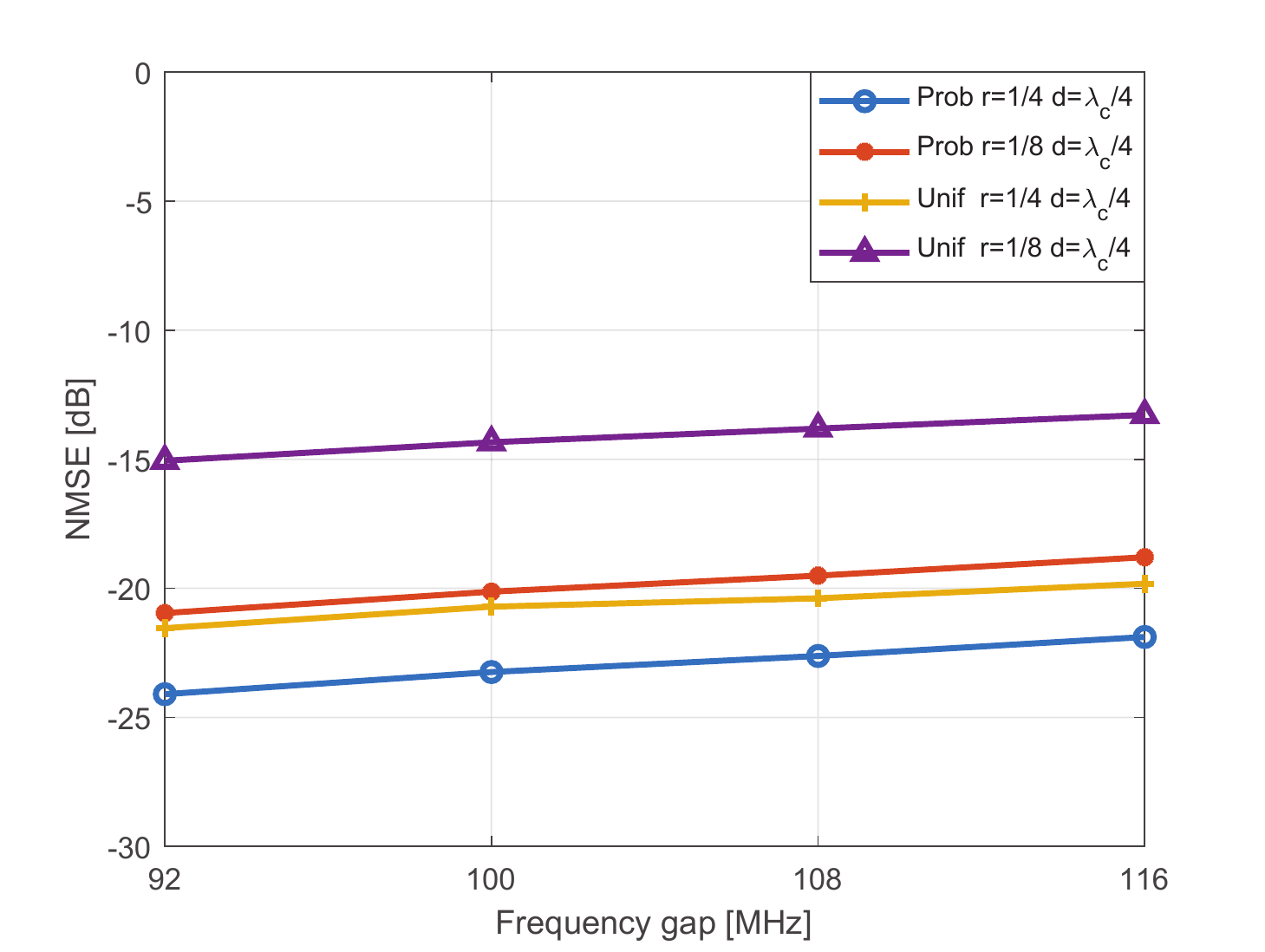}
	\caption{The NMSEs of channel extrapolation versus frequency gaps.}
	\label{re_frequency_gap}
\end{figure}

In Fig. \ref{re_frequency_gap}, we respectively extract the channels of 4 neighbouring subcarriers from $\{\mathbf H^a,\mathbf G^a\}$ and $\{\mathbf H,\mathbf G\}$ and further evaluate the NMSE performance of the channel extrapolation scheme with respect to the subcarrier frequency gaps between channel matrices, where $d=\frac{\lambda_c}{4}$ and 4 different frequency gaps are considered.
It can be found that as the subcarrier frequency gap increases, the NMSE gradually increases.
However, the performance impact is not large, which means that the proposed scheme can achieve a good extrapolation performance even with larger frequency gap.
Furthermore, with the same $r$, the performance of the proposed scheme with the probabilistic selection strategy is always better than that with the uniform selection strategy, which verifies its effectiveness.

\begin{figure}
	\centering
	\includegraphics[width=105mm]{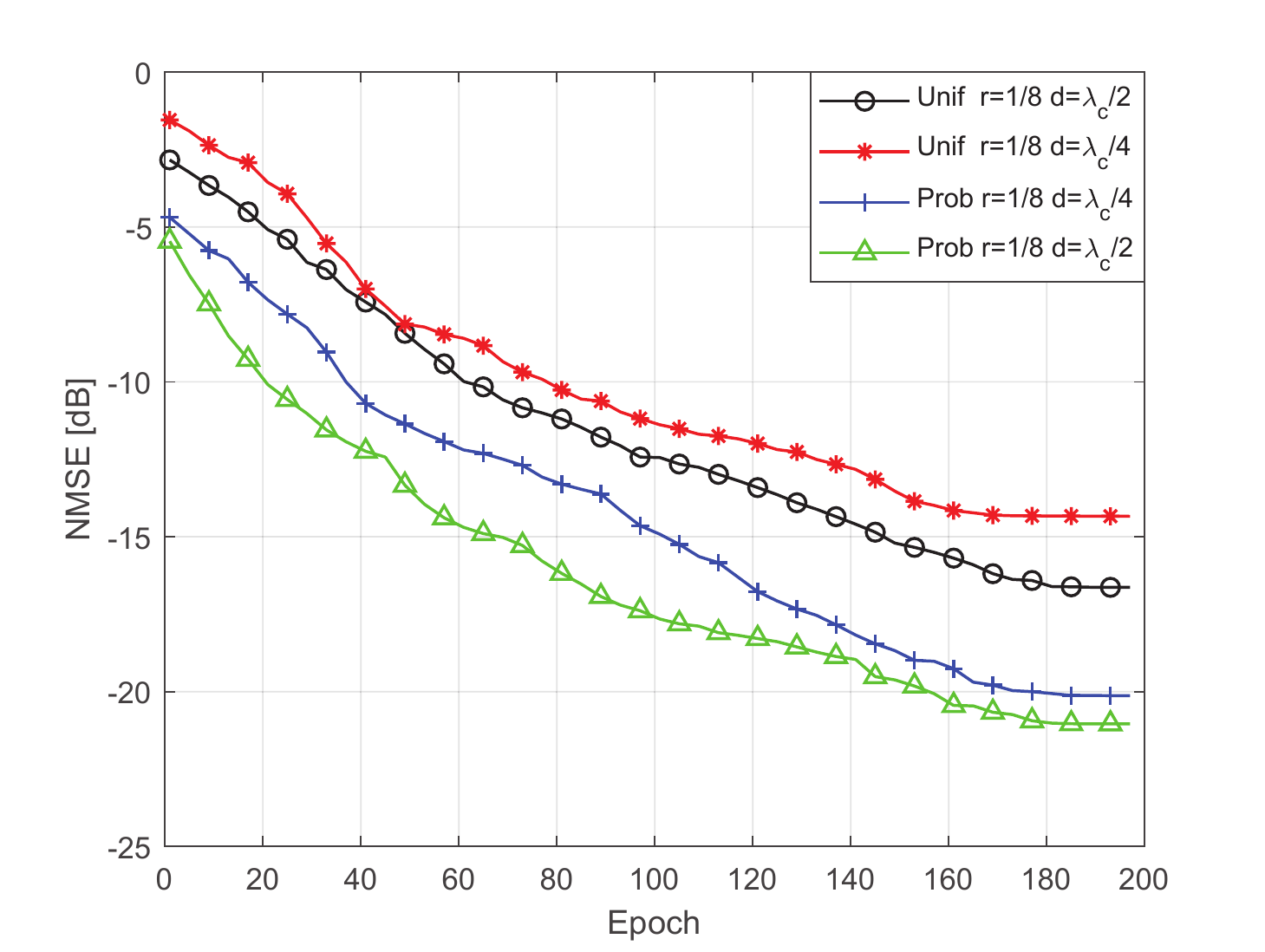}
	\caption{The NMSE of channel extrapolation versus epoches.}
	\label{re_Epoch}
\end{figure}

Fig. \ref{re_Epoch} studies the NMSE performance of the channel extrapolation scheme with probabilistic and uniform selection strategy versus epoch for network training, where $r=\frac{1}{8}$.
Obviously, it can be checked that the NMSE decreases with the epoch.
And it takes about 170 to 180 epoches to achieve the steady state, which proves the robustness of the proposed scheme.

\begin{figure}
	\centering
	\includegraphics[width=105mm]{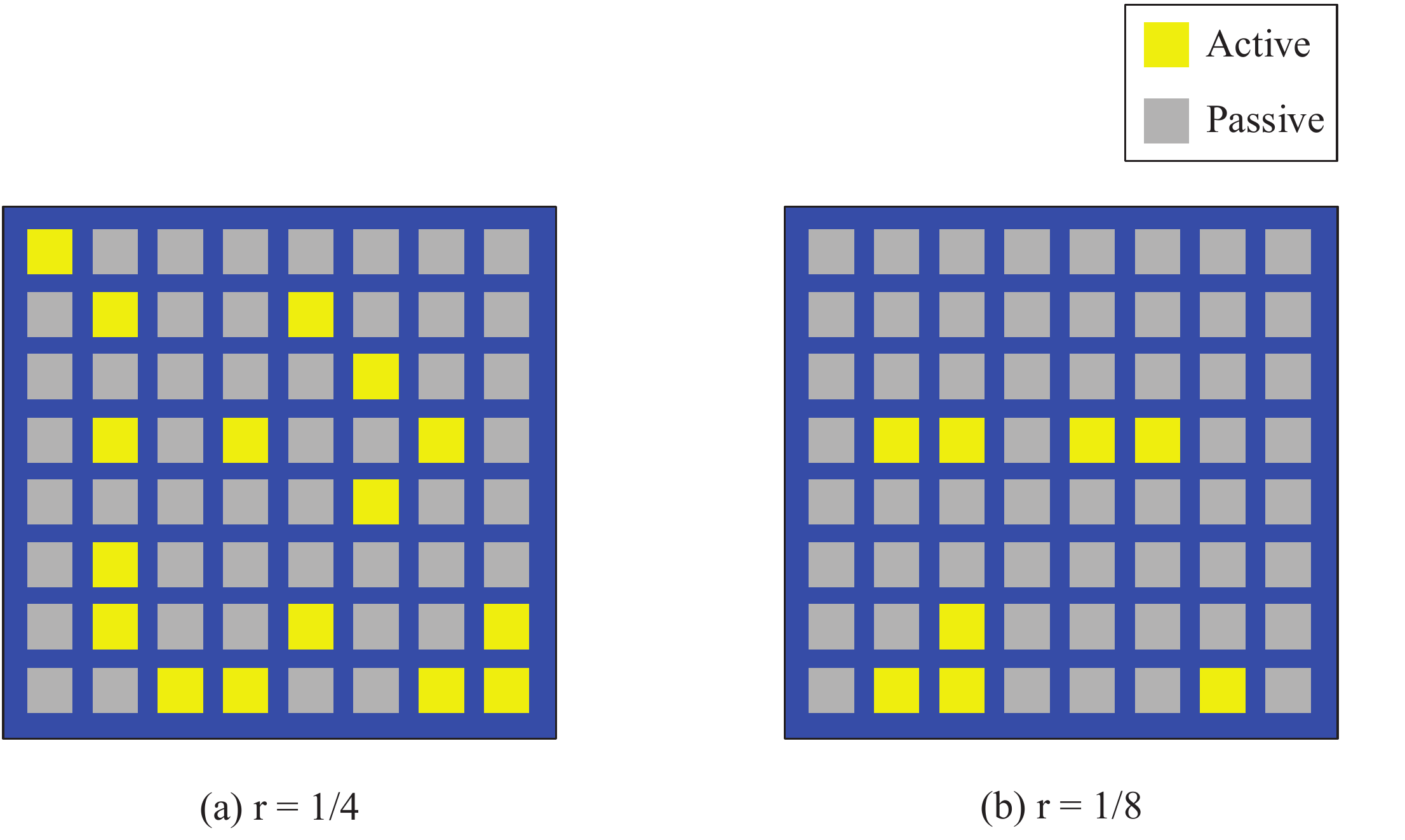}
	\caption{(a) The active antenna selection under $r=1/4$. (b) The active antenna selection under $r=1/8$.}
	\label{Pattern}
\end{figure}

Fig. \ref{Pattern} displays the active antenna selection results of $r=\frac{1}{4}$ and $r=\frac{1}{8}$, where $d=\frac{\lambda_c}{2}$. It can be found that there are 16 non-uniform active antennas selected under $r=\frac{1}{4}$ and 8 non-uniform active antennas selected under $r=\frac{1}{8}$.
Moreover, from the two sub-figures, it can be found that the uniform selection strategy is not optimal and the probabilistic selection strategy can achieve a better performance, which shows the effectiveness of the proposed probabilistic active antenna selection network.


\begin{figure}
	\centering
	\includegraphics[width=105mm]{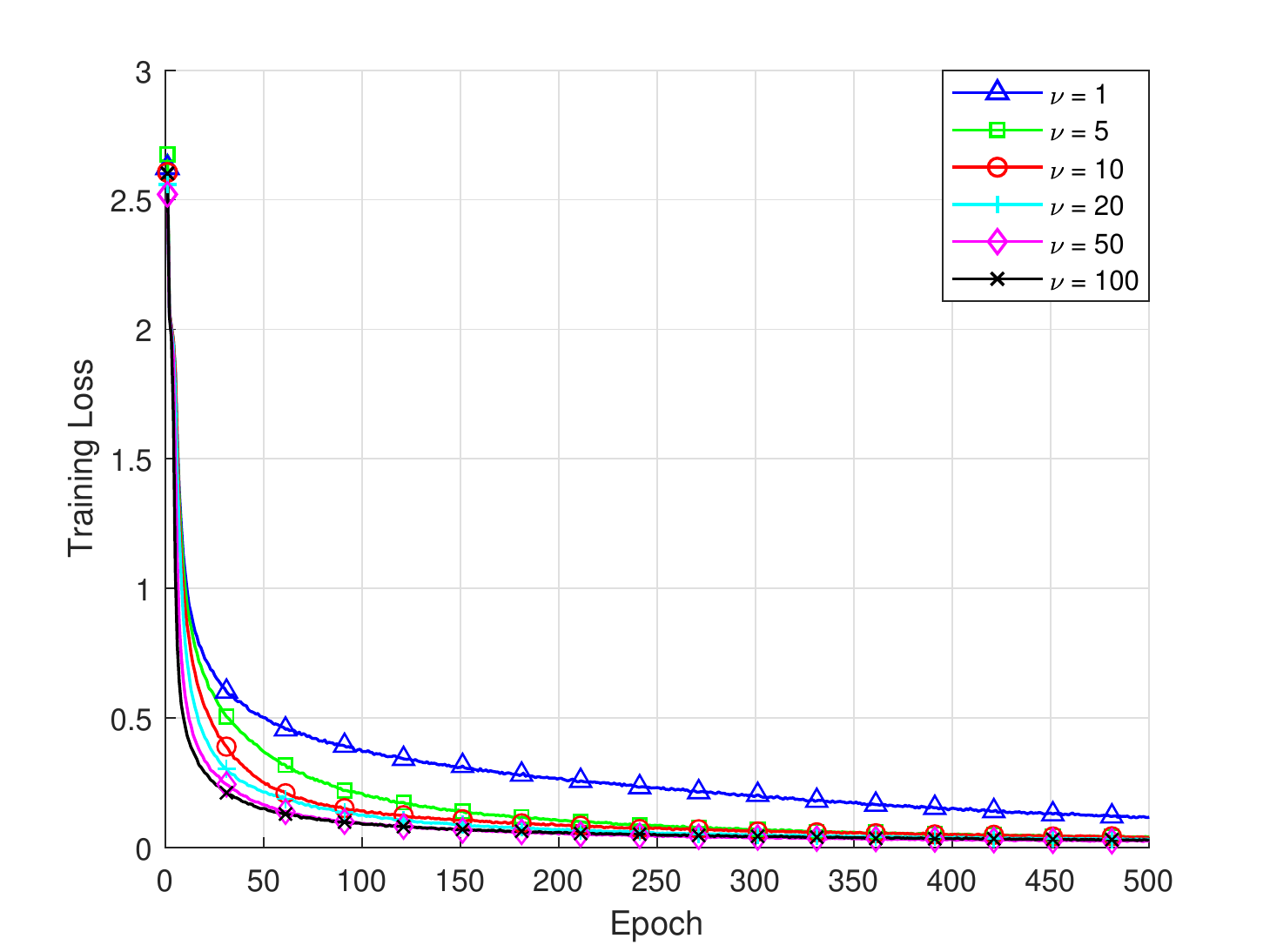}
	\caption{Training loss with different learning rate ratio for probabilistic selection.}
	\label{DPS_trainingloss_vs_Epoch_different_nu}
\end{figure}


Fig. \ref{DPS_trainingloss_vs_Epoch_different_nu} shows the training loss versus the epoch of the beam searching scheme with probabilistic selection strategy, where different learning rate ratio $\nu = \frac{\eta_\xi}{\eta_\omega}$ is considered and $r = \frac{1}{8}$.
It can be seen that the training loss decreases with the epoch.
Besides, it can be checked that the larger the learning rate ratio is, the lower the training loss will be.
This is because a larger learning rate ratio can accelerate the training of the selection network.
Moreover, we can see that the rate of convergence of the training loss is approaching a limit when $\nu = 100$.
It means that when $\nu > 100$, the increasing of $\nu$ has few benefits to the classification performance of the beam searching network.

\begin{figure}
	\centering
	\includegraphics[width=105mm]{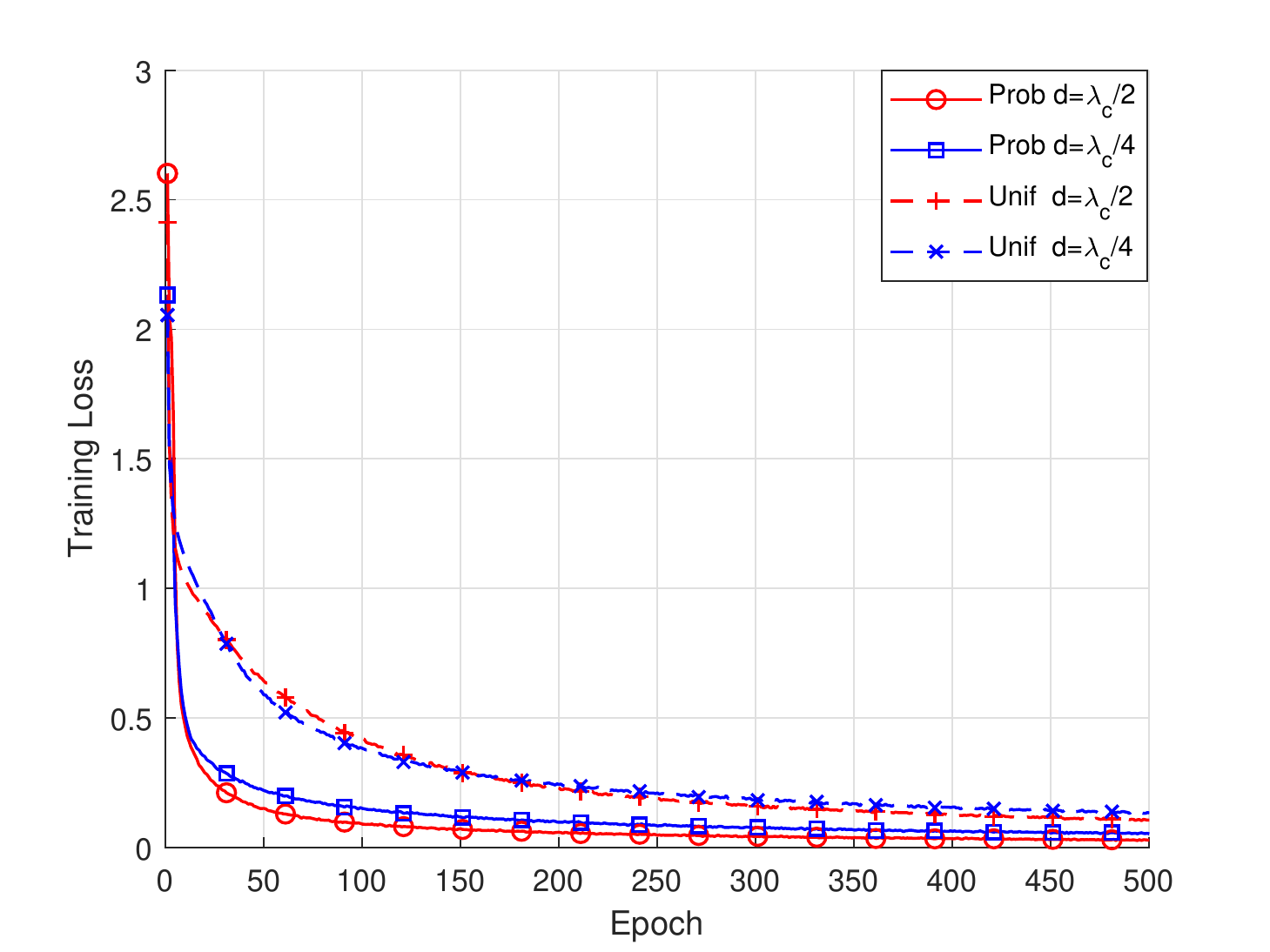}
	\caption{Training loss with different selection strategies and antenna spacings.}
	\label{Training_loss_vs_Epoch_different_sampway_and_space}
\end{figure}

In Fig. \ref{Training_loss_vs_Epoch_different_sampway_and_space}, we set $\nu = 100$ and $r = \frac{1}{8}$, with which a faster training loss convergence can be achieved as proved in Fig. \ref{DPS_trainingloss_vs_Epoch_different_nu}.
The figure shows the comparison of the training loss for the beam searching scheme with $d$ set as $\frac{\lambda_c}{2}$ and $\frac{\lambda_c}{4}$, respectively.
It can be seen that the performance with $d = \frac{\lambda_c}{2}$ is better than that with $d = \frac{\lambda_c}{4}$, which confirms the explanation provided for the results in Fig. \ref{re_compress_ratio}.
In addition, Fig. \ref{Training_loss_vs_Epoch_different_sampway_and_space} also shows the performance comparison between the uniform selection strategy and the probabilistic selection strategy, which verifies the considerable gain of the proposed probabilistic selection strategy.

\begin{figure}
	\centering
	\includegraphics[width=105mm]{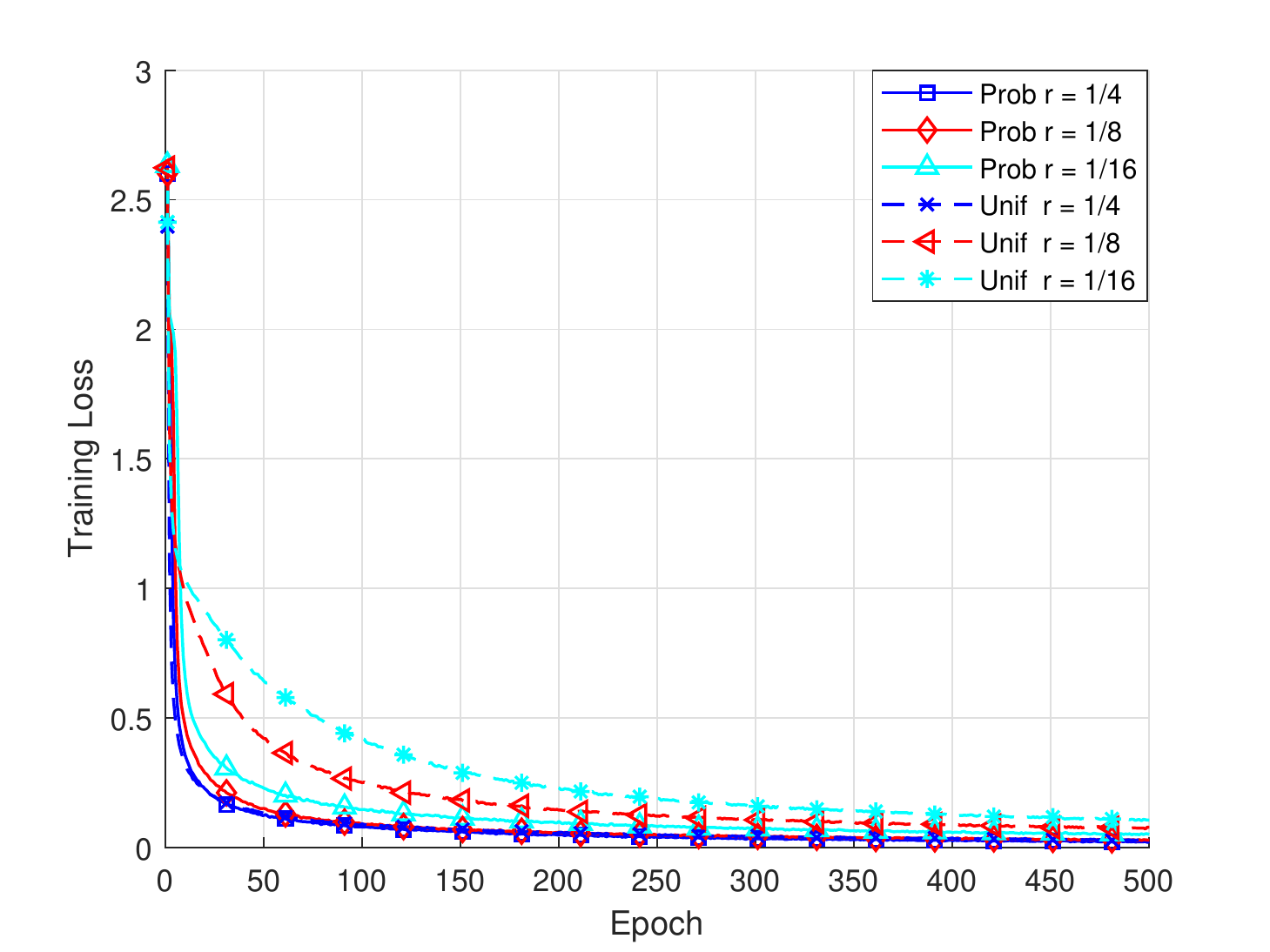}
	\caption{Training loss with different selection strategies and spatial compression ratios.} \label{Testloss_vs_Epoch_different_samplingrate_sampling_way}
\end{figure}

Fig. \ref{Testloss_vs_Epoch_different_samplingrate_sampling_way} shows the performance comparison between different $r$ for the beam searching scheme, where both probabilistic and uniform selection strategy are considered.
It can be checked that the performance enhances with the increase of $r$.
Besides, under each different $r$, the probabilistic selection strategy always provides a performance gain compared with the uniform selection strategy.
It is worth noting that when the spatial compression ratio becomes large enough, i.e., $r = \frac{1}{4}$, the probabilistic selection strategy shows a limited performance gain compared with the uniform selection strategy.
This is because that when increasing the number of sampled elements, the sampling strategy have less impact on the classification performance of the subsequent beam searching network.

\begin{figure}
	\centering
	\includegraphics[width=105mm]{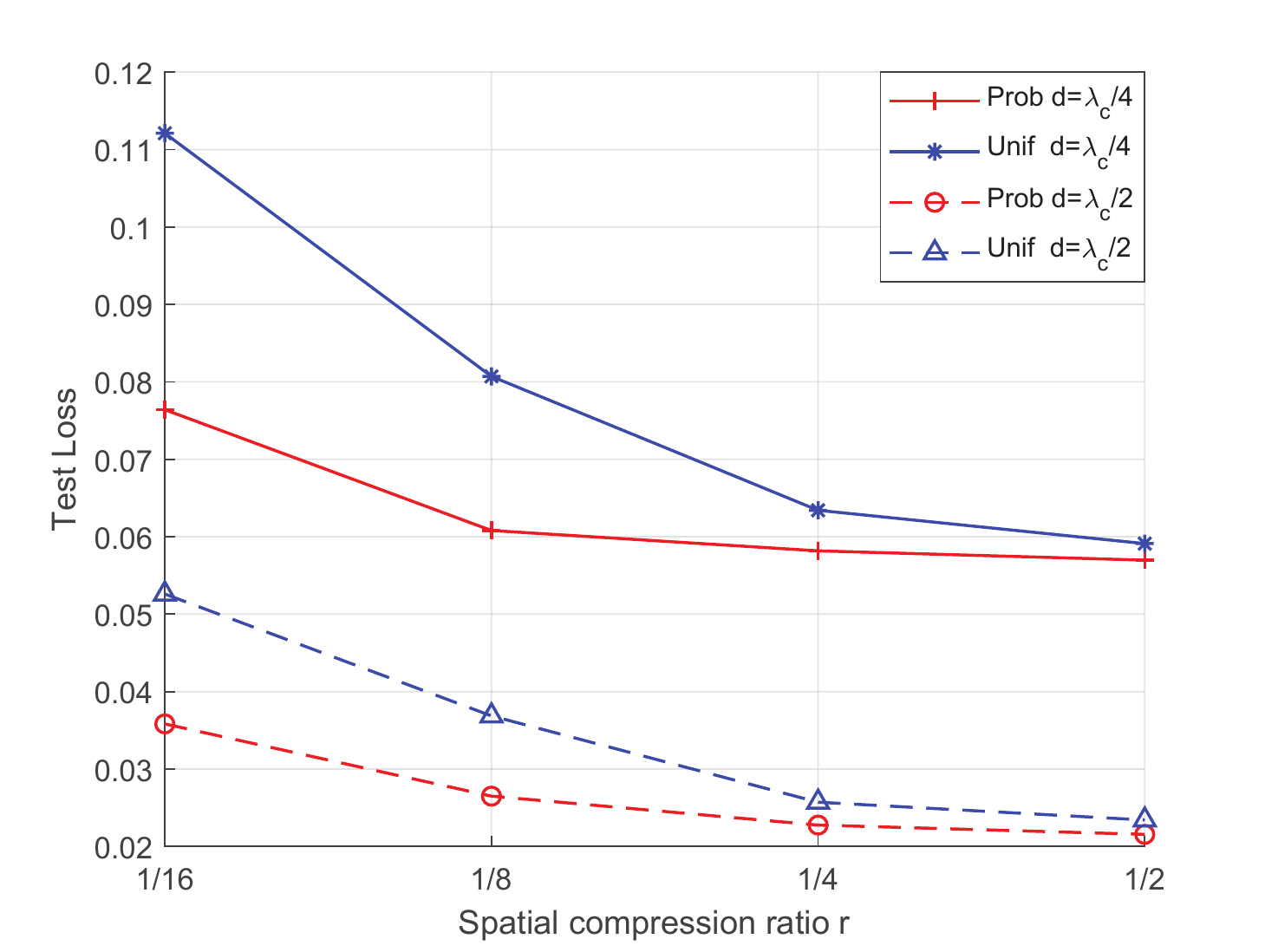}
	\caption{Test loss versus spatial compression ratio with different selection strategies and antenna spacings.}
	\label{Trainingloss_vs_samplingrate_different_samplingway_antennaspacing}
\end{figure}


In Fig. \ref{Trainingloss_vs_samplingrate_different_samplingway_antennaspacing}, we shows the classification performance on the test set versus the spatial compression ratio for the beam searching scheme, where both probabilistic and uniform selection strategy with two different antenna spacings are considered.
Results show that the test loss decreases with increasing $r$ for both selection strategy.
Moreover, Fig. \ref{Trainingloss_vs_samplingrate_different_samplingway_antennaspacing} illustrates the significant gain of the probabilistic selection strategy at a lower spatial compression ratio, i.e., $r = \frac{1}{16}$ and $r = \frac{1}{8}$, compared with the uniform selection strategy.
In addition, the gap between the test loss of probabilistic and uniform selection strategies reduces as $r$ increases.
Especially, when $r = \frac{1}{2}$, the test loss of probabilistic and uniform selection strategies are very close, for both $d = \frac{\lambda_c}{2}$ and $d = \frac{\lambda_c}{4}$, respectively, which has been explained in the description of Fig. \ref{Testloss_vs_Epoch_different_samplingrate_sampling_way}.

\begin{figure}
	\centering
	\includegraphics[width=105mm]{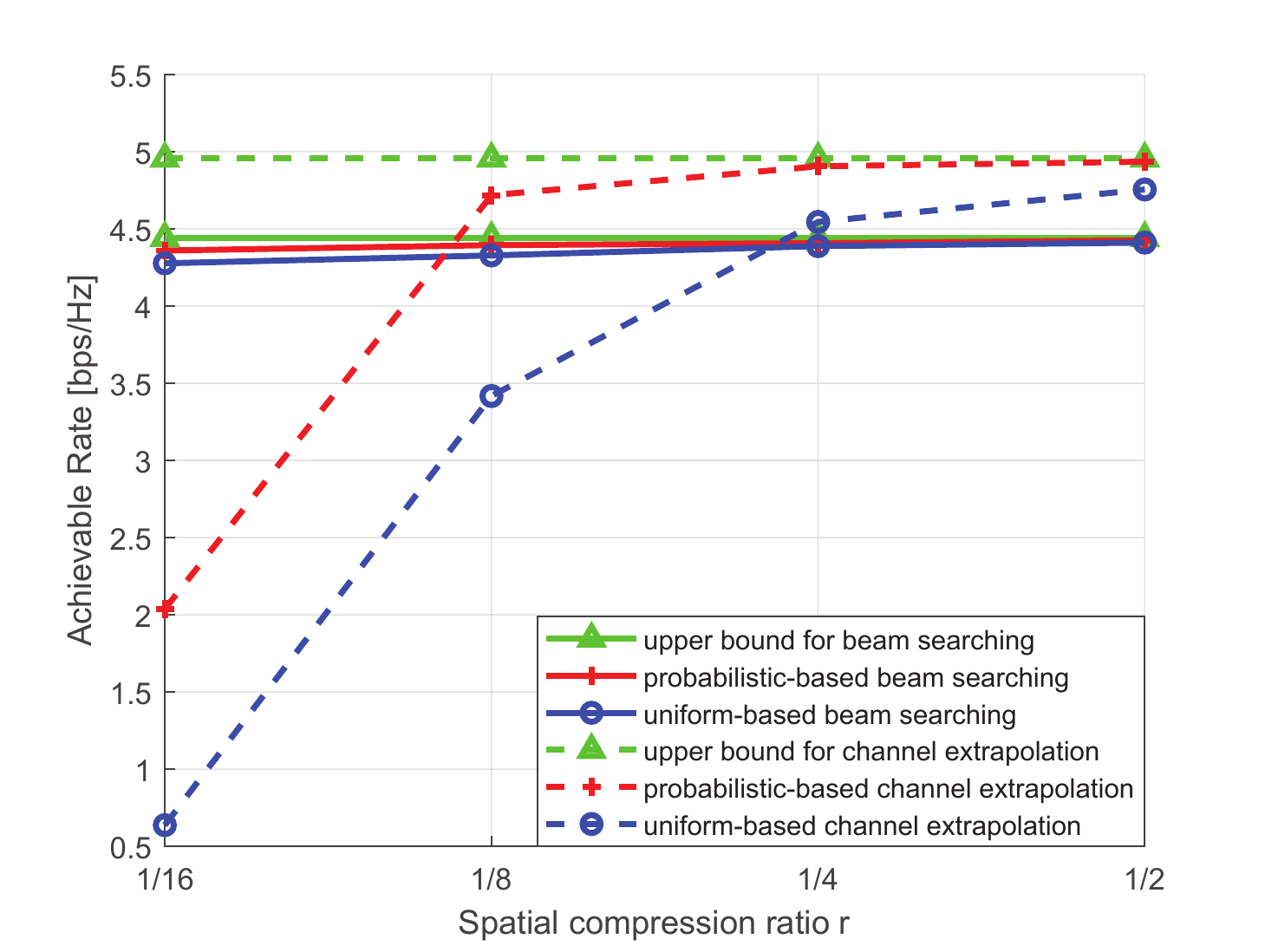}
	\caption{{Achievable rate versus spatial compression ratio with different selection strategies for the channel extrapolation scheme and the beam searching scheme, with $d=\frac{\lambda_c}{2}$}.}
	\label{rate}
\end{figure}

Fig. \ref{rate} shows the comparisons of the achievable rate for the two proposed schemes when $d=\frac{\lambda_c}{2}$ and the signal-to-noise ratio is 30 dB.
It can be seen that due to the limitation of the designed codebook $\mathcal B$, the upper bound of the achievable rate for the beam searching scheme is lower than that for the channel extrapolation scheme.
Moreover, since the optimal beamforming vector in \eqref{optimal_solution} highly depends on the performance of channel extrapolation, the achievable rate is lower for the channel extrapolation scheme compared with the beam searching scheme when $r=\frac{1}{16}$.
On the other hand, the achievable rate for the beam searching scheme is stable when $r$ reduces.
This demonstrates the robustness of the proposed beam searching scheme, which requires fewer active elements to obtain a considerable rate.
Furthermore, when $r$ increases, the achievable rates for the channel extrapolation scheme become higher than that for the beam searching scheme.
This can be explained that as the channel extrapolation performance greatly improved when increasing $r$, a better beamforming vector can be found based on \eqref{optimal_solution} and \eqref{beam_search_equ_pro} compared with the codebook $\mathcal B$.
For both the beam searching scheme and the channel extrapolation scheme, the achievable rates with the probabilistic selection strategy are higher than those with the uniform selection strategy, which again verifies the gain of the probabilistic sampling compared with the uniform sampling.

\section{Conclusions}
\label{conclusion}


In this paper, we have examined the active element-aided RIS communication system and proposed two DL-based schemes, i.e., the channel extrapolation scheme and the beam searching scheme, respectively.
For both schemes, the probabilistic sampling theory has been utilized to find the optimal locations of the active RIS elements.
Moreover, a CNN-based channel extrapolation network has been designed to extrapolated the full channels for data transmission from the estimated partial channels in the channel extrapolation scheme, whereas an FNN-based beam searching network has been designed to achieve the direct mapping from the estimated partial channels to the optimal beamforming vector for data transmission in the beam searching scheme.
The efficient BP was utilized to optimize the proposed networks during training.
Simulation results showed the effectiveness of the proposed DL-based schemes.

\begin{appendices}
\section{Feasibility of Channel Extrapolation}
\label{appendix_a}
From \eqref{frequency_domain_channel},
define the parameter set for $\mathbf H$ as $\mathcal{Q}_h(f_c) =\{h_{p, f_c}, \tau_{h,p}, \phi_{h,p}, \varphi_{h,p} \}_{p=1}^{P_h} $.
With the fixed structure of $\mathbb R$, $\mathbf H$ can be derived from $\mathcal{Q}_h(f_c)$.
Thus, $\mathcal{Q}_h(f_c) $ can be seen as the physical intrinsic factor of link along $\mathbb S\rightarrow\mathbb R$.
Before proceeding, we give the following definitions:

\textbf{Definition 1}: The mapping function $\bm \Phi_{\mathcal N,f_c}$ from the physical intrinsic factor set $\mathcal Q_h(f_c)$
to the channel $\mathbf H$ can be written as
\begin{align}
\label{p2c_mapping2_N}
\bm \Phi_{\mathcal N, f_{c}}: \{\mathcal{Q}_{h}(f_c)\} \to \{{\mathbf{H}}\},
\end{align}
where the sets $\{\mathcal{Q}_{h}(f_c)\}$ and $\{{\mathbf{H}}\}$ are the domain and codomain of $\bm \Phi_{\mathcal N, f_{c}}$, respectively.

Under fixed scattering scenario, if the number of RIS elements is large enough, we can extract $\mathcal  Q_{h}(f_c)$ from $\mathbf H$, which can be easily checked from \eqref{frequency_domain_channel}.
With the physical meanings of $\mathcal Q_{h}(f_c)$, we have the following bijective mapping\cite{channel_map} relation
\begin{align}
\mathcal Q_{h}(f_c)\leftrightarrow \mathbf H.
\end{align}

Thus, the above defined  mapping function \eqref{p2c_mapping2_N} is bijective, which means that $\mathcal Q_{h}(f_c)$ corresponds to one unique channel $\mathbf H$, and vice versa.
Then, the inverse mapping of $\bm \Phi_{\mathcal N, f_{c}}$ exists and  can be expressed as
\begin{align}
\label{c2p_mapping_N}
\bm \Phi_{\mathcal N, f_{c}}^{-1}: \{{\mathbf H}\} \to \{\mathcal{Q}_{h}(f_c)\}.
\end{align}

\textbf{Definition 2}: The mapping function $\bm \Phi_{\mathcal M, f_{c}}$ from $\mathcal Q_h(f_c)$
to the partial channel $\widetilde{\mathbf{H}}$ can be denoted as
\begin{align}
\label{p2c_mapping}
\bm \Phi_{\mathcal M, f_{\text{c}}}: \{\mathcal{Q}_{h}(f_c)\} \to \{\widetilde{\mathbf{H}}\},
\end{align}
where the sets $\{\mathcal{Q}_{h}(f_c)\}$ and $\{\widetilde{\mathbf{H}}\}$ are the domain and codomain of $\bm \Phi_{\mathcal M, f_{c}}$, respectively.

Since $\mathcal  M$ is a subset of $\mathcal N$ and $\widetilde {\mathbf H}$ is formed by the elements in
$\mathbf H$, with the results in {Definition 1}, it can be determined that the mapping function \eqref{p2c_mapping} is bijective
when the elements in $\mathcal M$ are sufficient.
Correspondingly,  the inverse mapping of $\bm \Phi_{\mathcal M, f_{c}}$ is
\begin{align}
\label{c2p_mapping}
\bm \Phi_{\mathcal M, f_{c}}^{-1}: \{\widetilde{\mathbf H}\} \to \{\mathcal{Q}_{h}(f_c)\}.
\end{align}

With the bijective properties of the mapping functions in \eqref{p2c_mapping2_N} and \eqref{p2c_mapping}, we can obtain the following proposition.

\textbf{Proposition 1} \cite{channel_map}: For the given communication environment
and RIS structure, the mapping relation from the partial channel $\widetilde{\mathbf H}$
to the full channel $\mathbf H$ can be characterized by  the function $\mathbf \Psi_{\mathcal M,f_{c} \to \mathcal N,f_{c}}$ defined as
\begin{align}
\label{a2d_mapping1}
\mathbf \Psi_{\mathcal M,f_{c} \to \mathcal N,f_{c}} = \mathbf \Phi_{\mathcal N,f_{c}} \circ \mathbf \Phi^{-1}_{\mathcal M,f_{c}}: \{\widetilde{\mathbf H} \} \to \{\mathbf H \},
\end{align}
where $(\cdot)\circ(\cdot)$ denotes the composite mapping operation.

The above proposition demonstrates that the extrapolation of $\mathbf H$ from $\widetilde{\mathbf H}$ is feasible.
We can consider a similar process
for $\mathbf G$  and $\widetilde{\mathbf G}$ and verify the feasibility of the proposed channel extrapolation \cite{yang_propo}.
Then, with feasible mapping between the partial channels and the full channels, we can effectively recover $\mathbf H$ and $\mathbf G$ from $\widetilde{\mathbf H}$ and $\widetilde{\mathbf G}$, respectively.

\section{Feasibility of Beam Searching}
\label{appendix_b}
Within the beam searching scheme, the optimal beamforming vector $\bm{\theta}^s$ at $\mathbb R$ is chosen from the codebook $\mathcal B$.
Since the feasible mapping from the partial channels $\widetilde{\mathbf H}$ and $\widetilde{\mathbf G}$ to the full channels $\mathbf H$ and $\mathbf G$ is existing.
With the extrapolated $\mathbf H$ and $\mathbf G$, we can traverse all the possible vectors in $\mathcal B$ and find the optimal beamforming vector $\boldsymbol\theta^s$ by utilizing (\ref{object}) as the performance metric.
Obviously, one set $\{\mathbf H,\mathbf G\}$ corresponds to only one determined $\boldsymbol\theta^s$.
Hence, there exists explicit mapping between $\{\widetilde{\mathbf H},\widetilde{\mathbf G}\}$ and $\boldsymbol\theta^s$ within $\mathcal B$.
Accordingly, the beam searching is also feasible. For clarity, we give
the following proposition.

\textbf{Proposition 2}: If $\boldsymbol\theta$ is selected from $\mathcal B$,
then there exists a specific mapping relation from $\{\widetilde{\mathbf H},\widetilde{\mathbf G}\}$
to the optimal $\boldsymbol\theta^s$.
Correspondingly, this mapping is expressed as
\begin{align}
\label{a2b_mapping}
\mathbf{\Pi}_{\mathcal M \to \mathcal B}
:
\left\{\widetilde{\mathbf H}, \widetilde{\mathbf G}\right\} \to
\left\{\boldsymbol\theta^s\right\}.
\end{align}

\end{appendices}

\linespread{1.3}
\balance

\end{document}